\documentclass[]{aa}
\usepackage{graphics}

\begin {document}
\title{The NASA Astrophysics Data System: Overview}

\thesaurus{04(04.01.1)}
\author{M. J. Kurtz\and G. Eichhorn\and A. Accomazzi\and C. Grant
\and S. S. Murray\and J. M. Watson}
\institute{Harvard-Smithsonian Center for Astrophysics, Cambridge, MA 02138}

\offprints{M. J. Kurtz}
\mail{M. J. Kurtz}

\date{Received / Accepted}

\titlerunning{ADS: Overview}
\authorrunning{M. J. Kurtz et al.}

\maketitle

\sloppy

\begin {abstract}

The NASA Astrophysics Data System Abstract Service has become a key
component of astronomical research.  It provides bibliographic
information daily, or near daily, to a majority of astronomical
researchers worldwide.

We describe the history of the development of the system and its
current status.  Urania (\cite{1996AAS...189.0603B}), and the ADS role
in the emerging electronic astronomical data environment are
discussed.  Astronomy is unique in that it already has a fully
functional data resource, where several of the most important data
sources exist on-line and inter-operate nearly seamlessly.  The ADS
and the Strasbourg Data Center (CDS; \cite{1998adass...7..470G}) form
the core of this resource.

We show several examples of how to use the ADS, and we show how ADS use
has increased as a function of time.  Currently it is still increasing
exponentially, with a doubling time for number of queries of 17
months.

Using the ADS logs we make the first detailed model of how scientific
journals are read as a function of time since publication.  We find
four distinct components.  We directly compare the readership rate
with the citation rate for scientific articles as a function of age.
Citations generally follow reads, but there are some differences.

The main journals of astronomy have differences in the ways they are
read and cited.  We discuss these from a number of different aspects.

The impact of the ADS on astronomy can be calculated after making some
simple assumptions.  We find that the ADS increases the efficiency of
astronomical research by 333 Full Time Equivalent (2000 hour) research
years per year, and that the value of the early development of the ADS for
astronomy, compared with waiting for mature technologies to be
adopted, is 2332 FTE research years.

A full technical description of the ADS is in three companion
articles: \cite{gei}, \cite{aa}, and \cite{csg}.  The ADS is available
at http://adswww.harvard.edu/.

\end{abstract}

\section {\label {intro} Introduction}

The NASA Astrophysics Data System Abstract Service (hereafter ADS,
except in section \ref {history}) is now a central facility of
bibliographic research in astronomy.  In a typical month (March
1999) it is used by more than 20,000 individuals, who make $\sim$
580,000 queries, retrieve $\sim$ 10,000,000 bibliographic entries, read
$\sim$ 400,000 abstracts and $\sim$ 110,000 articles, consisting of
$\sim$ 1,100,000 pages.  The ADS is a key element in the emerging
digital information resource for astronomy, which has been dubbed
Urania (\cite{1996AAS...189.0603B}).  The ADS is tightly
interconnected with the major journals of astronomy, and the major
data centers.

The present paper serves as an introduction to the system, a
description of its history, current status, use, capabilities, and
goals.  Detailed descriptions of the ADS system are in the
companion papers: The design and use of the search engine is in
\cite {gei}; hereafter SEARCH.  The architecture, indexing system, and
mirror maintenance is in \cite {aa}; hereafter ARCHITECTURE.  Finally
the methods we use to maintain and update the data base, and to
maintain communication with our collaborating data centers and
journals (primarily via bibcodes, \cite{1995VA.....39R.272S}) is in
\cite {csg}; hereafter DATA.

In section \ref {history} we discuss the history of the ADS, paying
particular note of the persons and events which were most important to
its development.  Section \ref {status} briefly discusses the current
status of the system, the data it contains, and the hardware,
software, and organizational methods we use to maintain and distribute
these data.  Urania, and especially the ADS role in it, is discussed
in section \ref {Urania}.  The current capabilities and use of the
system are shown in section \ref {use}; with section \ref {examples}
showing example queries, and section \ref {stats} showing how ADS use
has changed over time.  In section \ref{journals} we show how current
use varies as a function of the age of an article and the journal it
was published in; in \ref{read-use} we develop a multi-component model
which accurately describes the whole pattern of article use as a
function of age; in \ref{cite-use} we compare the similarities and
differences of readership information with citation histories; in
\ref{journal-use} we examine several aspects of the readership pattern
for the major journals.  Finally, in \ref{impact}, we estimate the
impact of the ADS on astronomy.

\section {\label {history} Historical Introduction}

The ADS Abstract Service had its beginnings at the conference
Astronomy from Large Data-bases, held in Garching in 1987.  There
\cite {1988alds.proc..143A} discussed the desirability of building a
natural language interface to a set of astronomical abstracts
(Astronomy and Astrophysics Abstracts (A\&AA) was the model) using
software from Information Access Systems, Inc. (IAS; E. Busch was the
president of IAS).  \cite {1988alds.proc..453W} discussed the existing
abstract services.  At this meeting G. Shaw (who was representing IAS)
saw the paper by \cite {1988alds.proc..113K}, and noticed that the
vector space classification methods developed by M.\ J.\ Kurtz for the
numerical classification of stellar spectra were very similar to those
developed by P.\ G.\ \cite{1966MBR.....1..479O} for the classification
(and thus natural language indexing) of text. Ossorio's methods were
the basis of the proposal by \cite {1988alds.proc..143A}; Ossorio was
the founder of IAS.  Shaw suggested Ossorio and Kurtz meet.  Also at
this conference \cite{1988alds.proc..489S} presented the NASA plan for
an astrophysics data system, and Shaw met G.\ Squibb.

This meeting of Kurtz and Ossorio took place in January 1988, in
Boulder, CO.  By the end of the meeting it was clear that the
technical difficulties involved in creating an abstract service with a
natural language index could be overcome, if the data could become
available.  A preliminary mathematical solution to the problem was
developed, under the assumption that A\&AA would be the source of the
abstracts.  This technique was later called the ``statistical Factor
Space'', factor analysis being one of the tools used to create the
vector space.

Over the next year NASA moved to implement the \cite
{1988alds.proc..489S} plan for the establishment of a network based,
distributed system for access and management of NASA astrophysics data
holdings, the Astrophysics Data System.  Shaw and Ossorio founded a
new company, Ellery Systems, Inc., which obtained the systems
integration contract for the ADS.  During this time Shaw, Ossorio,
Kurtz, and S.S.  Murray all spoke often about the abstract service as
an integral part of the emerging ADS system, and the abstract service,
and Factor Space, became nearly synonymous with the ADS project.  No
actual work was done to implement the abstract service during this
time, Ossorio and Kurtz worked on applying their vector space
classification techniques to galaxy morphologies (\cite
{1989daa..conf..121O}, \cite {1990PaReL..11..507K}), while Murray used
the original, non-statistical, Factor Space methods of \cite
{1966MBR.....1..479O} to build a small ($\sim$40 documents) natural
language indexing system for demonstration purposes.

During the next three years the ADS was built
(\cite{1992adass...1...35G}), but without a literature retrieval
service, which was listed as a future development.  No NASA funds were
devoted to the abstract service during this time.  Independently Kurtz
and Watson set out to obtain the data necessary to build a prototype
system; keyword data was received from the IAU (International
Astronomical Union) Thesaurus project (\cite{1992PASAu..10..134S},
\cite{1993asth.book.....S}), and from the NASA Scientific and
Technical Information (STI) branch (\cite{1990GIQ.....7..123P}).  A
breakthrough occurred in mid 1990 when the Astronomische Rechen
Institut graciously provided Watson with magnetic tape copies of the
two 1989 volumes of Astronomy and Astrophysics Abstracts.  By the end
of 1990 Kurtz (1991, 1992) had built a prototype abstract retrieval
system, based on the statistical Factor Space.

In April, 1991 F. Giovane and C. Pilachowski organized a meeting near
Washington, D.C. on ``On-Line Literature in Astronomy.''  At this
meeting \cite {boy91} discussed the desire of the American
Astronomical Society (AAS) to publish on-line journals, \cite {kur91}
discussed the prototype system, and pointed out the types of queries
which would be made possible if a natural language abstract system
were combined with the Strasbourg Data Centers's (CDS) SIMBAD
(\cite{1988alds.proc..323E}) database and with the Institute for
Scientific Information's Science Citation Index
(\cite{1979cita.book.....G}), and \cite{van91} discussed the desire of
the National Space Science Data Center (NSSDC) to create a database of
scanned bitmaps of journal articles.  Also at this meeting were
representatives of NASA's STI branch, who indicated that they would be
willing to provide the abstracts from the STI (often called NASA
RECON) abstracts database (\cite{1990GIQ.....7..149W}).

Near the end of the meeting \cite{mur91} outlined the possibilities
inherent in the previous talks.  He described a networked data system
where a natural language query system for the STI abstracts would work
jointly with the CDS/SIMBAD object name index to point astronomers to
relevant abstracts, article bitmaps, and electronic journal articles.
Save that the World Wide Web (\cite{ber94}) has taken the place of the
proprietary network software created for the ADS project by Ellery
Systems Inc., and that the ADS has taken over responsibility for the
bitmaps from the NSSDC, the current system is essentially identical to
the one predicted by \cite {mur91}.

Following the meeting the NSSDC group (\cite{1993adass...2..137W})
organized the STELAR project, which held a series of meetings where
many of the issues involved in electronic journals were discussed, and
a consensus was reached on allowed uses of copyrighted journal article
bitmaps.

In the spring of 1992 Murray took over the direct management of
the ADS project; G. Eichhorn was hired as project manager.  The
decision was made to proceed forthwith with the development of an
abstract service based on the STI abstracts.  Because the STI abstract
system is differently structured than the A\&AA system the statistical
Factor Space was abandoned in favor of a more traditional entropy
matching technique (\cite{sal83}, see SEARCH).

The new system was working with a static database by fall, and was
shown at the Astronomical Data Analysis Software and Systems II
meeting in Boston (\cite{1993adass...2..132K}).  The production system
was released in February 1993, as part of the package of ADS services,
still part of the proprietary ADS network system.  Abstract Service
use quickly became more than half of all ADS use.

By summer 1993 a connection had been made between the ADS and SIMBAD,
permitting users to combine natural language subject matter queries
with astronomical object name queries (\cite{1994AAS...184.2802G}).
This connection was enabled by the use of the bibcode (see DATA).  We
believe this is the first time an internet connection was made to
permit the routine, simultaneous, real-time interrogation of
transatlanticly separated scientific databases.

By early 1994 The World Wide Web (\cite{ber94}) had matured to where
it was possible to make the ADS Abstract Service available via a web
forms interface; this was released in February.  Within five weeks of
the initial WWW release use of the Abstract Service quadrupled (from
400 to 1600 users per month).

By the end of 1994 the ADS project had again been restructured,
leaving primarily the WWW based Abstract Service as its principle
service.  Also the STELAR project at NSSDC ended, and the ADS took
over responsibility for creating the database of bitmaps.

The first full article bitmaps, which were of Astrophysical Journal
Letters articles, were put on-line in December 1994
(\cite{1994AAS...185.4104E}).  By the summer of 1995 the bitmaps were
current and complete going back ten years.  At that time the
Electronic ApJ Letters (\cite{1995AAS...187.3801B}) went on-line.
>From the start the ADS indexed the EApJL, and pointed to the
electronic version.  Also from the beginning the reference section of
the EApJL pointed (via WWW hyperlinks) to the ADS abstracts for
articles referenced in the articles; again this was enabled by the use
of the bibcode.

Also during this time the NASA STI branch became unable to provide
abstracts of the journal articles in astronomy.  In order to continue
the abstract service cooperative arrangements were made with nearly
every astronomical research journal, as well as a number of other
sources of bibliographic information.  DATA describes these
arrangements in detail.  

The next year (1996) saw nearly every astronomy journal which had not
already joined into collaboration with ADS join.  Also in 1996 the
American Astronomical Society purchased the right to use a subset of
the Science Citation Index, and gave these data to ADS
(\cite{1996AAS...189.0607K}).

\section{\label{status} The Current System}

Currently the ADS system consists of four semi-autonomous (to the
user) abstract services covering Astronomy, Instrumentation, Physics,
and Astronomy Preprints.  Combined there are nearly 1.5 million
abstracts and bibliographic references in the system.  The Astronomy
Service is by far the most advanced, and accounts for $\sim 85$\%\ of
all ADS use; it ought be noted, however, that the Instrumentation
Service contains more abstracts than Astronomy, and a subset of that
service is used by the Society of Photo-Optical Industrial Engineers
as the basis of the official SPIE publications web site.

All of what follows will refer only to the Astronomy service.

\subsection{Data}

Here is a brief overview of the data in the ADS system, a complete
description is in DATA.

\subsubsection{Abstracts}

The ADS began with the abstracts from the NASA STI database, in
printed form these abstracts were the union of the International
Aerospace Abstracts and the NASA Scientific and Technical Abstracts and
Reports (NASA STAR).  While the STI branch has had to substantially
cut back on their abstracting of the journal literature, we still get
abstracts of NASA reports and other materials from them.

We now receive basic bibliographic information (title, author, page
number) from essentially every journal of astronomy.  Most also send
us abstracts, and some cannot send abstracts, but allow us to scan
their journals, and we build abstracts through optical character
recognition.  Finally we receive some abstracts from the editors of
conference proceedings, and from individual authors.

The are $\sim$500,000 different astronomy articles indexed in the ADS,
the database is nearly complete for the major journals articles
beginning in 1975.

\subsubsection{Bitmaps}

The ADS has obtained permission to scan, and make freely available
on-line, page images of the back issues of nearly all of the major
journals of astronomy.  In most cases the bitmaps of current articles
are put on-line after a waiting period, to protect the financial
integrity of the journal.  DATA describes the current status of these
efforts.

We plan to provide for each collaborating journal, in perpetuity, a
database of page images (bitmaps) from volume 1 page 1 to the first
issue which the journal considers to be fully on-line as published.
This will provide (along with the indexing and the more recent
archives held by the journals) a complete electronic digital library
of the major literature in astronomy.  

On a longer term we plan to scan old observatory reports, and defunct
journals, to finally have a full historical collection on-line.  This
work is beginning with a collaboration with the Harvard Preservation
Project (\cite{1997AAS...191.3502E}; \cite{1995VA.....39..161C}).

\subsubsection{Links}

ADS responds to a query with a list of references and a set of
hyperlinks showing what data is available for each reference.  There
are $\sim$1.73 million hyperlinks in the ADS, of which
$\sim$ 31\%\ are to sources external to the ADS project.

The largest number of external links are to SIMBAD, NED, and the
electronic journals.  A rapidly growing number, although still small
in comparison to the others, are to data tables created by the
journals and maintained by the CDS and the ADC at Goddard.  SEARCH
describes the system of hyperlinks in detail.

\subsubsection{Citations and References}

The use of citation histories is a well known and effective tool for
academic research (\cite{1979cita.book.....G}); their inclusion in the
ADS has been planned since the conception of the service.  In 1996 the
AAS purchased a subset of the Science Citation Index from the
Institute for Scientific Information, to be used in the ADS; this was
updated in 1998.  This subset only contains references which were
already in the ADS, thus it is seriously incomplete in referring to
articles in the non-astronomical literature.  This citation
information currently spans January 1982-September 1998.

The electronic journals all have machine readable, web accessible,
reference pages.  The ADS points to these with a hyperlink where
possible.  Several publishers allow us to use these to maintain
citation histories; we do this using our reference resolver software
(see ARCHITECTURE).  The same software is also used by some publishers
to check the validity of their references, pre-publication.

Additionally we use optical character recognition
to create reference and citation lists for the historical literature,
after it is scanned (\cite{1999AAS...195.8209D}).

\subsubsection{Collaboration with CDS/SIMBAD}

The Strasbourg Data Center (CDS) has long maintained several of the
most important data services for astronomy
(e.g. \cite{1971BICDS...1....2J}; \cite{1973BICDS...4...27J};
\cite{1998adass...7..470G}); access to parts of the CDS data via ADS
is a key feature of the ADS.

ADS users are able to make joint queries of the ADS bibliographic
database and the CDS/SIMBAD bibliographic data base.  When SIMBAD
contains information on a object which is referred to in a paper whose
reference is returned by ADS then ADS also returns a pointer to the
SIMBAD data.  When a paper has a data table which is kept on-line at
the CDS the ADS returns a pointer to it.  The CDS-ADS collaboration is
at the heart of Urania (section \ref{Urania}).  More recently ADS has
entered into a collaboration with the National Extragalactic Database
(NED; \cite{1988alds.proc..335H}, \cite{1992adass...1...47M}) which is
similar to the SIMBAD portion of the CDS-ADS collaboration.

\subsection {Search Engine}

The basic design assumption behind the search engine, and other user
interfaces, is that the user is an expert astronomer.  This differs
from the majority of information retrieval systems, which assume that
the user is a librarian.  The default behavior of the system is to
return more relevant information, rather than just the most relevant
information, assuming that the user can easily separate the wheat from
the chaff.  In the language of information retrieval this is favoring
recall over precision.  SEARCH describes the user interface in detail.

\subsection{Hardware and Software Architecture}

The goals of our hardware and software systems are speed of
information delivery to the user, and ease of maintainability for the
staff.  We thus pre-compute many things during our long indexing
process for later use by the search engine; we have highly optimized
all code which is run by user processes; we have developed a worldwide
network of mirror sites to speed up internet access.  ARCHITECTURE
describes these systems.

\subsection {Data Ingest}

The basic rule for what books and periodicals the ADS covers is: if it is
in the Center for Astrophysics library it should be in the ADS.  As a goal
we are still some ways from realization.  We have recently adopted a
second rule for inclusion: if it is referenced by an article in a
major scholarly journal of astronomy it should be in the ADS.  DATA
describes the ADS coverage, and ingest procedures.

\section{\label{Urania} Urania}

The idea that the internet could be used to link sources of
astronomical information into a unified environment is more than a
decade old; it was fully expressed in the planning for the old ADS
(\cite{1988alds.proc..489S}) and ESIS (\cite{1988alds.proc..137A})
projects.  These early attempts were highly data oriented, their
initial goals were the interoperability of different distributed data
archives, primarily of space mission data.

Astronomical data is highly heterogeneous and complex; essentially
every instrument has its quirks, and these must be known and dealt
with to reduce and analyze the data.  This quirky nature of our data
essentially prevented the establishment of standardized tools for data
access across data archives.

The new, hyperlink connected network data system for astronomy is
based on the highest level of data abstraction, object names and
bibliographic articles, rather than the lowest, the actual observed
data in archives.  This change in the level of abstraction has
permitted the creation of a system of extraordinary power.  This new
system, still unique amongst the sciences, has been dubbed Urania
(\cite{1996AAS...189.0603B}), for the muse of astronomy.

Conceptually the core of Urania is a distributed cross-indexed list
which maintains a concordance of data available at different sites.
The ADS maintains a list of sites which provide data organized on an
article basis for every bibliographic entry in the ADS database.  The
CDS maintains a list of articles and positions on the sky for every
object in the SIMBAD database.  The CDS also provides a name to object
resolver.  The possibility for synergy in combining these two data
systems is obvious; they have functioned jointly since 1993.

Surrounding this core, and tightly integrated with it, are many of the
most important data resources in astronomy, including the ADS Abstract
Service, SIMBAD, the fully electronic journals (currently ApJL, ApJ,
ApJS, A\&A, A\&AS, AJ, PASP, MNRAS, New Astronomy, Nature, and
Science), NED, CDS-Vizier, Goddard-ADC, and the ADS Article Service.
All these groups actively exchange information with the Urania core,
they point their users to it via hyperlinks, and they are pointed to
by it.

The astronomy journals which are not yet fully electronic, in that
they do not support hyperlinked access to the Urania core, also
interact with the system.  Typically they provide access to page
images of the journal, either through PDF files, or bitmaps from the
ADS Article Service, or both.  Bibliographic information is routinely
supplied to the ADS, and the SIMBAD librarians routinely include the
articles (along with those of the electronic journals) in the SIMBAD
object-article concordance.

While most data archives are not closely connected to the Urania
system there are some exceptions.  For example the National Center for
Supercomputing Application's Astronomy Digital Image Library
(\cite{1996adass...5..581P}) connects with the ADS bibliographical
data via links which are papers written about the data in the archive.
SIMBAD connects with the High Energy Astrophysics Science Archive
Research Center (HEASARC) (\cite{1992adass...1...52W}) archive using
the position of an object as a search key, HEASARC has an interface
which permits several archives to be simultaneously queried
\cite{1998adass...7..481M}, and a new data mining initiative between
CDS and the European Southern Observatory (ESO)
(\cite{1999adass...8..379O}) will connect the Vizier tables with the
ESO archives.  Several archives use the SIMBAD (and in some cases NED)
name resolver to permit users to use object name as a proxy for
position on the sky, the Space Telescope Science Institute (STScI)
Digital Sky Survey (\cite{1996stsc.rept.....P}) would be an example.
The Space Telescope-European Coordinating Facility archive
(\cite{mur95}) allows ADS queries using the observing proposals as
natural language queries, and the Principal Investigator names as
authors.

The establishment and maintenance of the Urania core represents a
substantial fraction of the ADS service.  SEARCH discusses the user
interface to the set of hyperlinks, ARCHITECTURE discusses the methods
and procedures we use to implement and maintain the links, and DATA
discusses the data sharing arrangements we have with other groups, and
presents a complete listing of all our data sources.

\section{\label{use} Capabilities, Usage Patterns, and Statistics}

\subsection{\label{examples} Examples}

The ADS answers about 5,000,000 queries per year, covering a wide
range of possible query type, from the simplest (and most popular):
``give me all the papers written by (some author),'' to complex
combinations of natural language described subject matter and
bibliometric information.  Each query is essentially the sum of
simultaneous queries (e.g. an author query and a title query), where
the evidence is combined to give a final relevance ranking
(e.g. \cite{1995InPrM..31..431B}).

The ADS once supported index term (keyword) queries, but does not
currently.  This is due to the incompatibility of the old STI
(\cite{1988NASAS7069.....N}) keyword system with the keywords assigned
by the journals (\cite {1990ApJ...357....1A};
\cite{1992A+A...253..A12.}; \cite{1992MNRAS.259......}) .  Work is
underway to build a transformation between the two systems
(\cite{1999adass...8..287L}; \cite{1999sigirconf..198L}).

Here we show four examples of simple, but sophisticated queries, to
give an indication of what is possible using the system.  A detailed
description of available query options is in SEARCH.  We encourage the
reader to perform these queries now, to see how the passage of time
has changed the results.

\begin{figure}
\resizebox{\hsize}{!}{\includegraphics{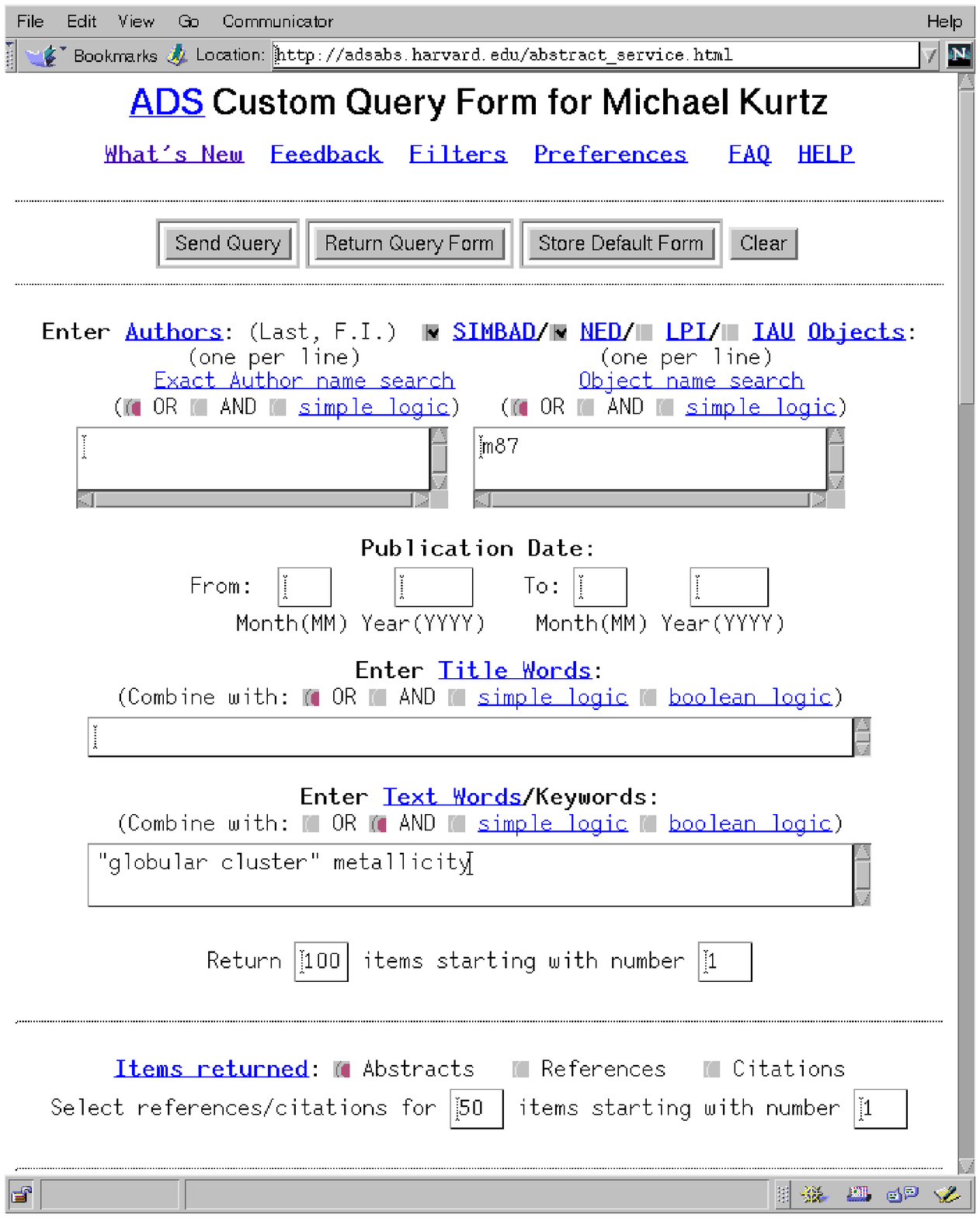}}
\caption[]{A query to the ADS Abstract Service      requesting a listing of papers on the metallicity of M87 globular      clusters.  SIMBAD, NED, the ADS phrase index, the ADS word index      and the ADS synonym list are all queried, the results are      combined and the list shown in figure \ref{m87.out} is      returned. }
\label{m87.query}
\end{figure}

Figure \ref{m87.query} shows how to make the query ``what papers are
about the metallicity of M87 globular clusters?''  This was the first
joint query made after the SIMBAD-ADS connection was completed in
1993.  

There are 1,765 papers on M87 in SIMBAD, NED, or both; there are 6,425
papers which contain the phrase ``globular cluster'' in ADS, and there
are 25,599 papers in ADS containing ``metallicity'' or a synonym
(abundance is an example of a synonym for metallicity).  The result,
which comes in a couple of seconds, is a list of just those 58 papers
desired.

\begin{figure}
\resizebox{\hsize}{!}{\includegraphics{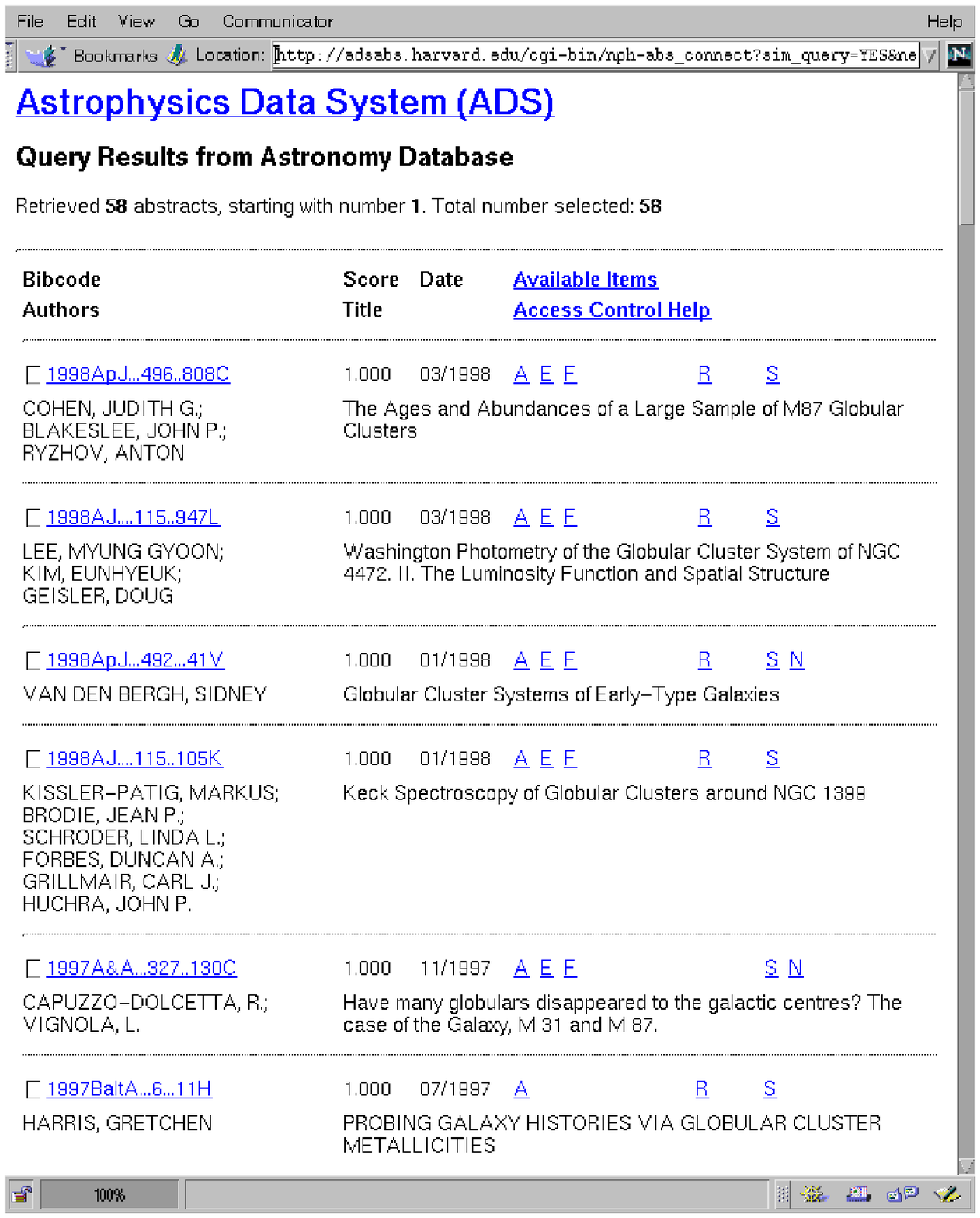}}
\caption[]{The top of the list ADS returns when the      query shown in figure \ref{m87.query} is made. }
\label{m87.out}
\end{figure}

Five different indices are mixed in this query: the SIMBAD
object---bibcode index query on M87 is logically OR'd with the NED
object---refcode index query for M87.  The ADS phrase index query for
``globular cluster'' is (following the user's request) logically AND'd
with the ADS word index query on metallicity, where metallicity is
replaced by its group of synonyms from the ADS astronomy synonym list
(this replacement is under user control).  If the user requires a
perfect match, then the combination of these simultaneous queries
yields the list of 58 papers shown in figure \ref{m87.out}.  Before
the establishment of the Urania core queries like this were nearly
impossible.

Another simple, but very powerful method for making ADS queries is to
use the ``Find Similar Abstracts'' feature.  Essentially this is an
extension of the ability to make natural language queries, whereby the
user can choose one or more abstracts to become the natural language
query.  This can be especially useful when one wants to read in depth
on a subject, but only knows one or two authors or papers in the
field.  This is a typical situation for many researchers, but
especially for students.

\begin{figure}
\resizebox{\hsize}{!}{\includegraphics{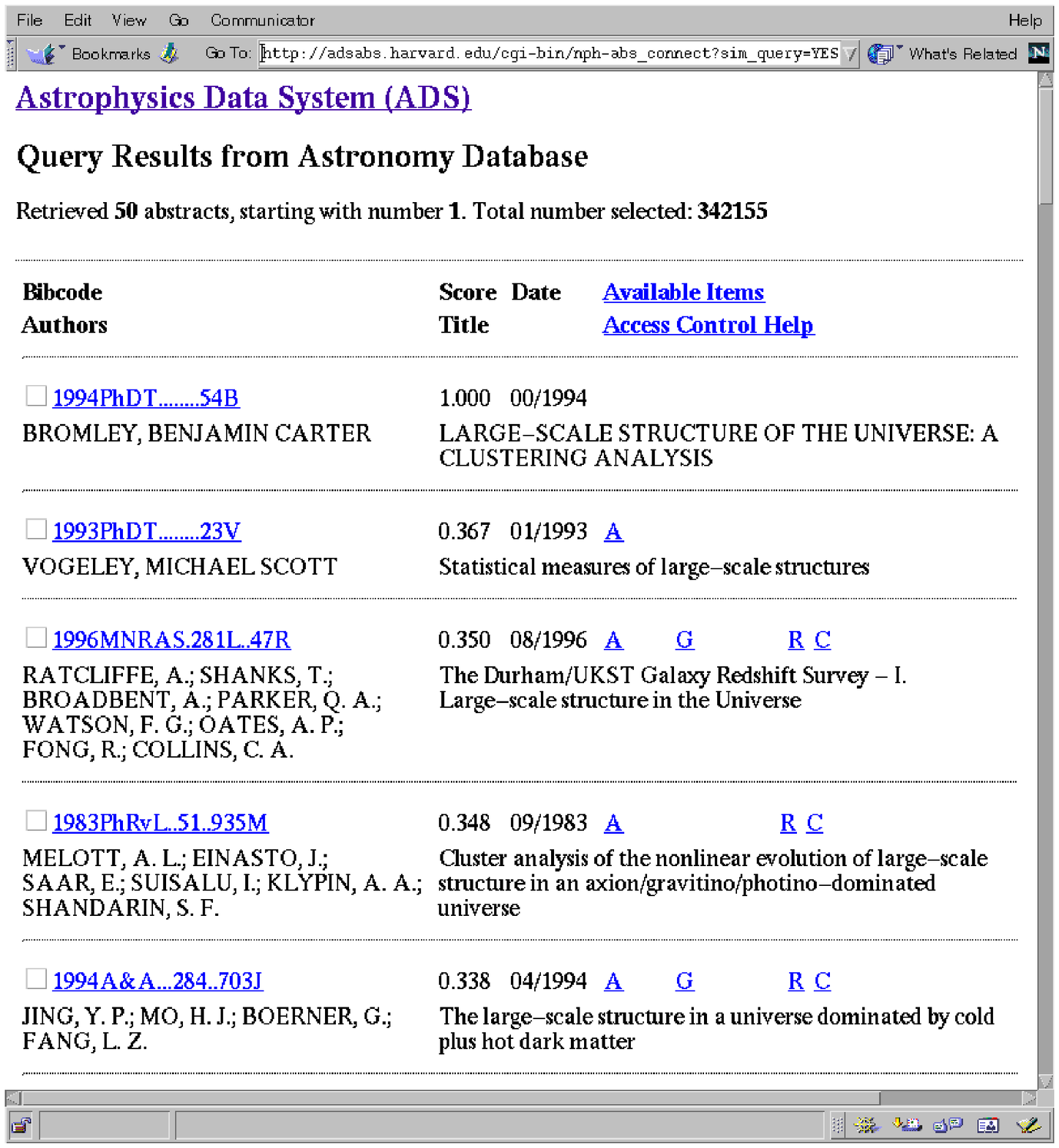}}
\caption[]{The top of the list of papers returned      when Ben Bromley's (1994) thesis is used as the      query. }
\label{bromley.list}
\end{figure}

As an example, suppose one is interested in Ben Bromley's (1994) PhD
thesis work.  Making an author query on ``Bromley'' gets a list of his
papers, including his thesis.  Next one calls up the abstract of the
thesis, goes to the bottom of the page, where the ``Find Similar
Abstracts'' feature is found, and clicks the ``Send'' button.  Figure
\ref{bromley.list} shows the top of the list returned as a result.
These are papers listed in order of similarity to Bromley's (1994)
thesis; note that the thesis itself is on top, as it matches itself
perfectly.  This list is a detailed subject matter selected custom
bibliography.

\begin{figure}
\resizebox{\hsize}{!}{\includegraphics{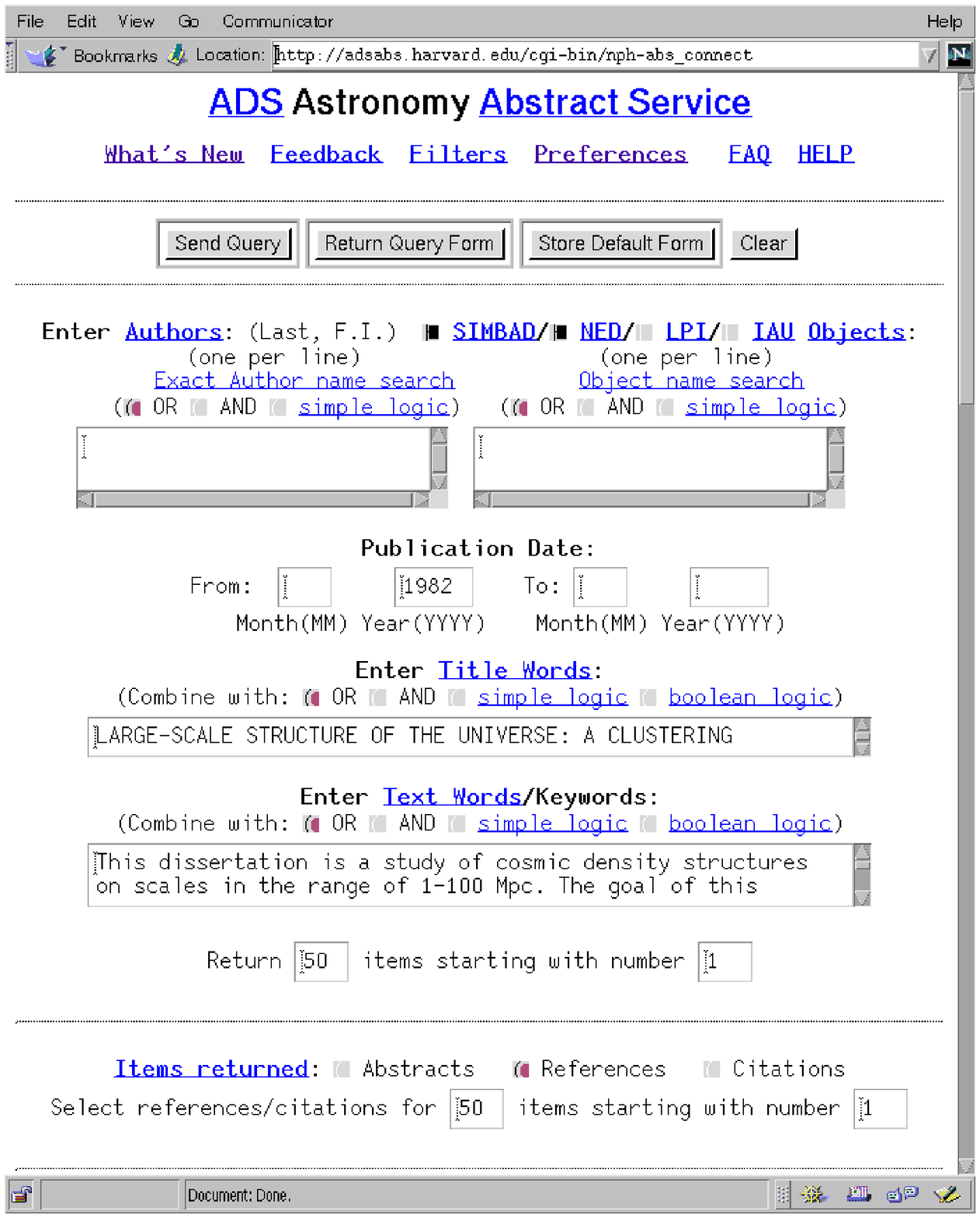}}
\caption[]{A query which returns the papers most      cited by the 50 papers most like Ben Bromley's (1994)      thesis. }
\label{bromley.query}
\end{figure}

\begin{figure}
\resizebox{\hsize}{!}{\includegraphics{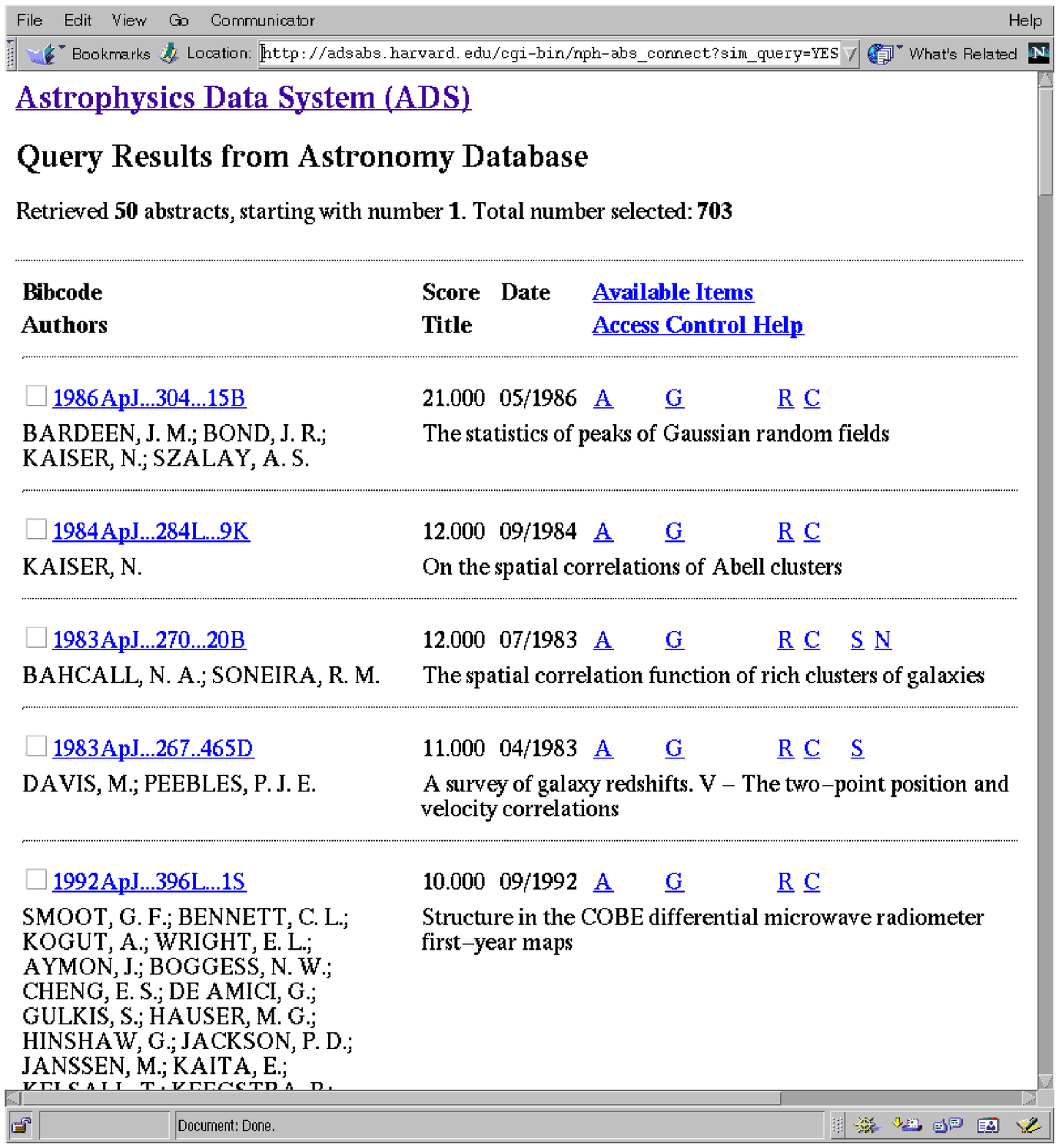}}
\caption[]{The top of the list of papers returned by      the query in Figure \ref{bromley.query}; these are the most cited      papers in a user defined very narrow      subfield. }
\label{bromley.refs}
\end{figure}

As a third example of ADS use figure \ref{bromley.query} shows an
intermediate step from the previous example (obtained by clicking on
the ``Return Query Form'' button, replacing the default ``Return Query
Results'' in the ``Find Similar Abstracts'' query.  Here we make one
change from the default setting: we change ``Items returned'' from the
default ``Abstracts'' to ``References.''  The result, shown in figure
\ref{bromley.refs} lists all the papers which are referenced in the 50
papers most like \cite{1994PhDT........54B}, sorted by the number of
times they appear in the 50 reference lists.  Thus the paper by
\cite{1986ApJ...304...15B} appears in 21 reference lists out of 50,
the paper by \cite{1983ApJ...267..465D} appears in 11 lists out of 50,
etc.  By this means one has a list of the most cited papers within a
very narrowly defined subfield specific to one's personal interest.  We
are not aware of any other system which currently allows this
capability.

\begin{figure}
\resizebox{\hsize}{!}{\includegraphics{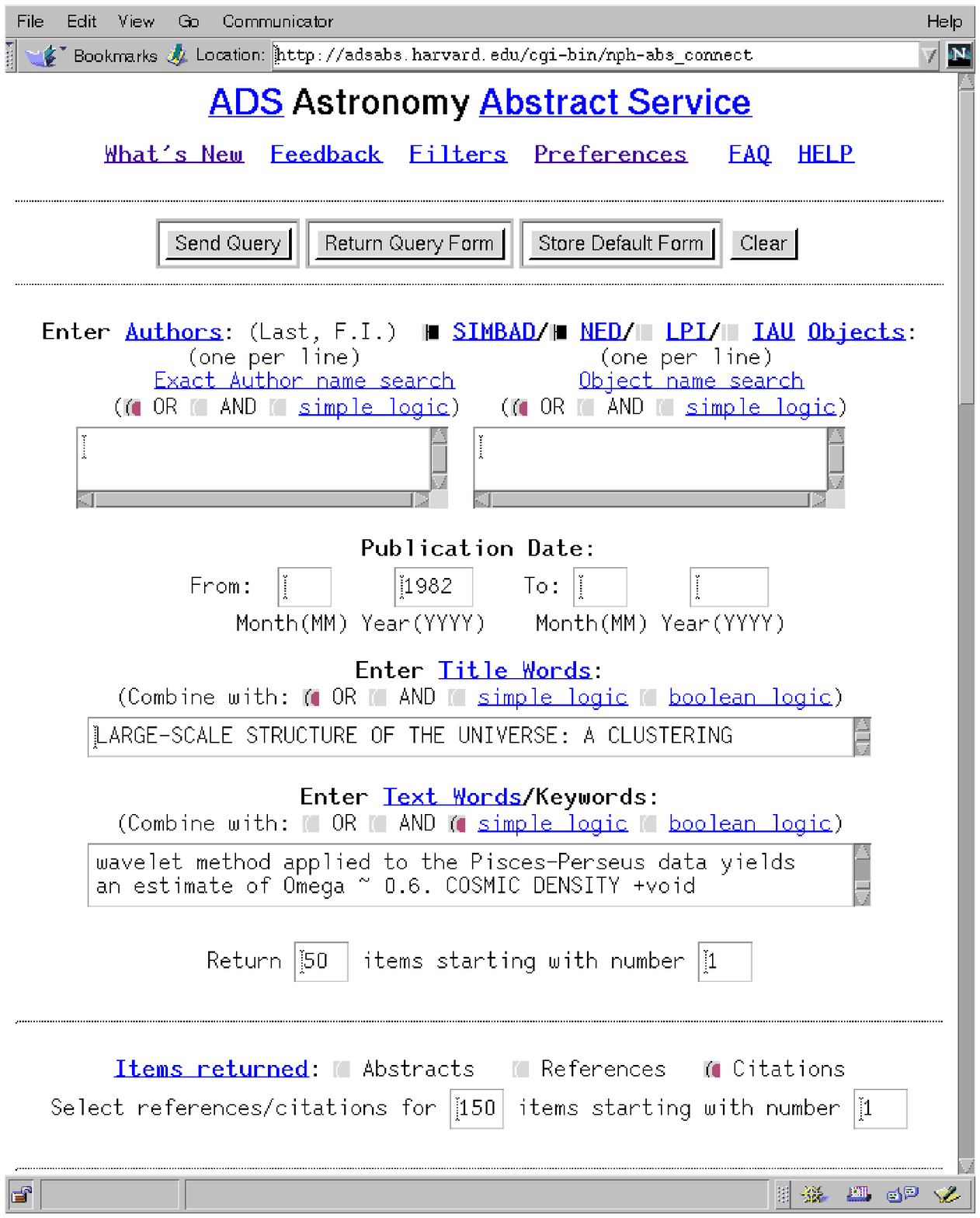}}
\caption[]{A query which returns the papers which      most cite the 150 papers most like Ben Bromley's (1994) thesis,      as modified by the requirement that they contain the word      ``void.'' }
\label{bromley.query2}
\end{figure}

\begin{figure}
\resizebox{\hsize}{!}{\includegraphics{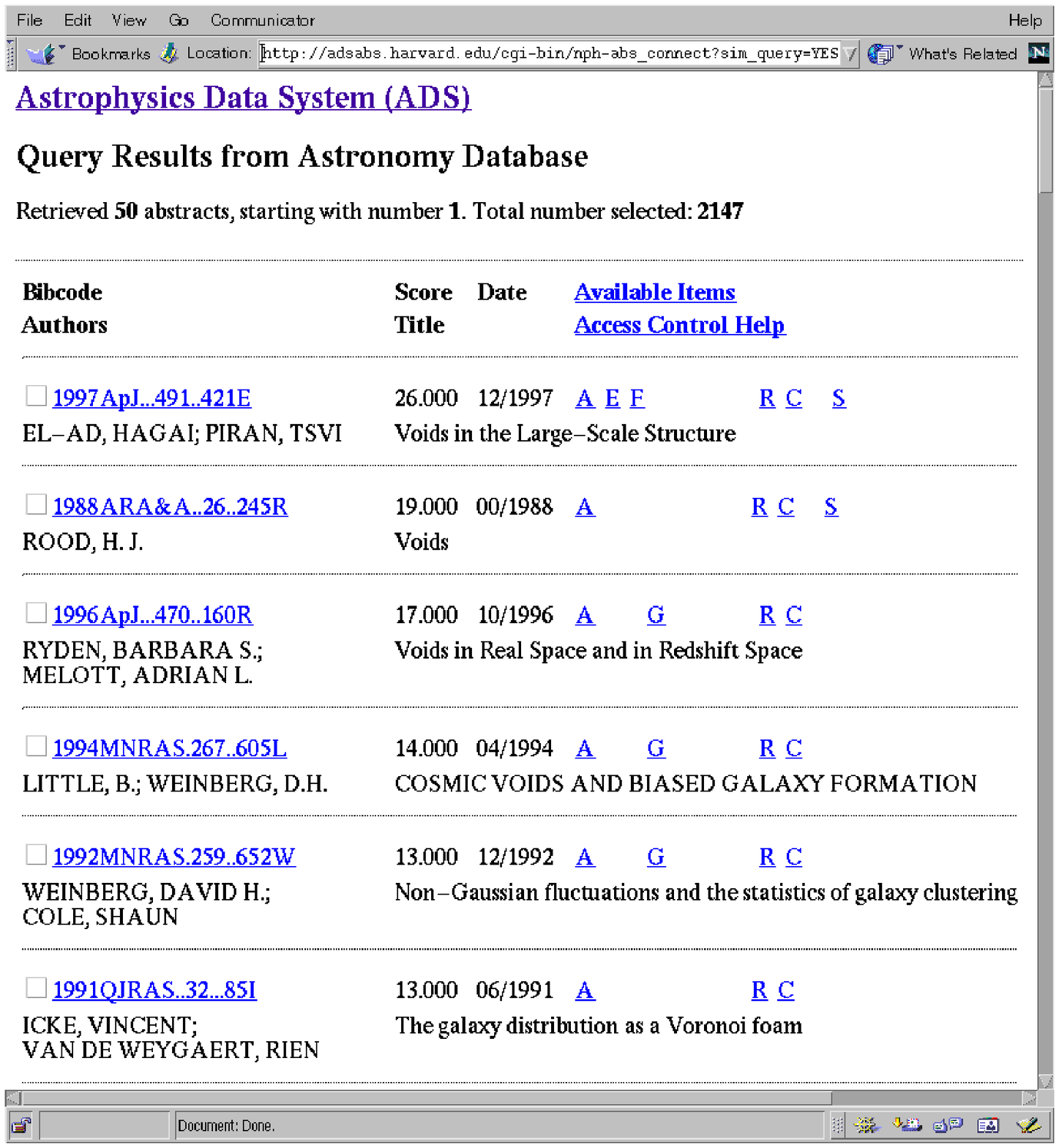}}
\caption[]{The top of the list of papers returned      by the query in Figure \ref{bromley.query2}; these are the papers      with the most extensive discussions of a user defined very narrow      subfield. }
\label{bromley.cites}
\end{figure}

Finally we show a somewhat more complex query in figure
\ref{bromley.query2}.  Here we modify the basic query (Bromley's
(1994) thesis) by requiring that the papers contain the word ``void.''
We do this by changing the logic on the text query to ``simple logic''
and adding ``+void'' to the query.  The returned papers to this query
would be very similar to those shown in figure \ref{bromley.list}, but
with all papers which do not contain the word ``void'' removed.  In
addition we change ``Items returned'' to be ``Citations,'' and
increase the number of papers to get the citations for to the top 150
closest matches to the query.  The result, shown in figure
\ref{bromley.cites}, are those papers which most cite the 150 papers
most like Bromley's (1994) thesis, modified by the requirement that
they contain the word ``void.''  Thus the paper by
\cite{1997ApJ...491..421E} cited 26 papers out of the 150, the paper
by \cite{1988ARA&A..26..245R} cited 19, etc.  These are the papers
with the most extensive discussions of a user defined very narrow
subfield.  This feature also is unique to the ADS.

\subsection{\label{stats} Use of the System}

In September 1998 ADS users made 440,000 queries, and
received 8,000,000 bibliographic references, 75,000 full-text
articles, and 275,000 abstracts (130,000 were individually selected,
the rest were obtained through a bulk retrieval process, which
typically retrieves between one and fifty), as well as citation
histories, links to data, and links to other data centers.  Of the
75,000 full-text articles accessed through the ADS in September 1998,
already 33\%\ were via pointers to the electronic journals.  This
number increased to 52\%\ in March 1999.

ADS users access and print (either to the screen, or to paper) more
actual pages than are printed in the press runs of all but the very
largest journals of astronomy.  In September 1998, 472,621 page images
were downloaded from the ADS archive of scanned bitmaps.  About 75\%\
of these were sent directly to a printer, 22\%\ were viewed on the
computer screen, and 2\%\ were downloaded into files; FAXing and
viewing thumbnail images make up the rest.  If the electronic journals
provide ``pages'' of information at the same rate as the ADS archive,
per article accessed (slightly more than 10 pages/article accessed),
then more than 750,000 ``pages'' were ``printed,'' on demand, in
September 1998 by ADS users.  This is about three times the number of
physical pages published in September 1998 by the {\it
PASP}.

Viewed as an electronic library the ADS, five years after its
inception, provides bibliographic information and services similar to
those provided by the sum of all the astronomy libraries in the world,
combined.  The Center for Astrophysics Library, an amalgamation of the
libraries of the Harvard College Observatory and the Smithsonian
Astrophysical Observatory, is one of the largest, most complete, and
best managed astronomy libraries in the world.  For several years the
CfA Library has been keeping records of the number of volumes
reshelved, as a proxy for the number of papers read (library users are
requested not to reshelve anything themselves).  This number has
remained steady in recent years, and was 1117 in September 1998
(D.J. Coletti \&\ E.M. Bashinka 1998, personal communication).  If the
CfA represents 2--3\%\ of the use of astronomy libraries, worldwide
(the CfA has slightly more than 350 PhDs, the AAS has about 6800
members, the IAU about 8500, CfA users made 2.4\%\ of ADS queries in
September 1998, 5.7\%\ of articles in the ADS Astronomy database with
1998 publication dates had at least one CfA author), and if other
astronomers use their libraries at the same rate as astronomers at the
CfA, then worldwide there would have been 37,000--56,000 reshelves in
September 1998.  In September 1998 ADS provided access to 75,000 full
text articles and 130,000 individually selected abstracts, as well as
substantial other information; current use of ADS is clearly similar
to the sum of all current traditional astronomy library use.

\begin{figure}
\resizebox{\hsize}{!}{\includegraphics{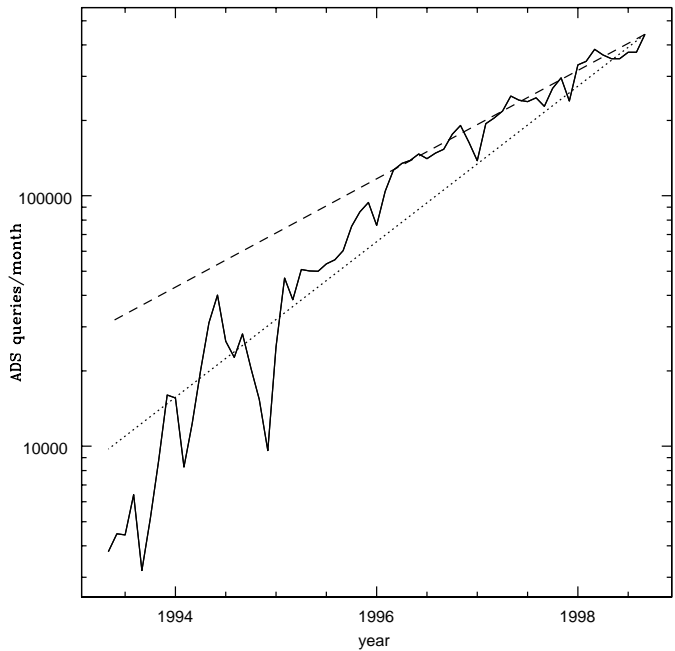}}
\caption[]{The number of queries made each month to      the ADS Abstract service.  The dotted line represents a yearly      doubling, while the dashed line represents a doubling period of      17 months, a reasonable match to the recent      data. }
\label{num.queries}
\end{figure}

ADS use continues to increase.  Figure \ref{num.queries} shows the
number of queries made each month to the ADS Abstract Service from
April 1993 to September 1998, the dotted straight line represents a
yearly doubling, which represents the five year history reasonably
well.  Since 1996 use has been increasing at a 17 month doubling rate,
shown by the dashed line in the figure.

It is difficult to determine the exact number of ADS users.  We track
usage by the number of unique ``cookies''\footnote{A cookie is a
unique identifier which WWW providers (in this case ADS) assign to
each user, and store on the users computer using the browser.} which
access ADS, and by the number of unique IP\footnote{Each Machine on
the internet has a unique IP (Internet Protocol) address.} addresses.
There are difficulties with each technique.  In addition many
non-astronomers find ADS through portal sites like Yahoo, which skews
the statistics.  In September 1998 10,000 unique cookies accessed the
full-text articles, 17,000 made queries, and 30,000 visited the site.
91\%\ of full-text users had cookies, but only 65\%\ of site visitors.

\begin{figure}
\resizebox{\hsize}{!}{\includegraphics{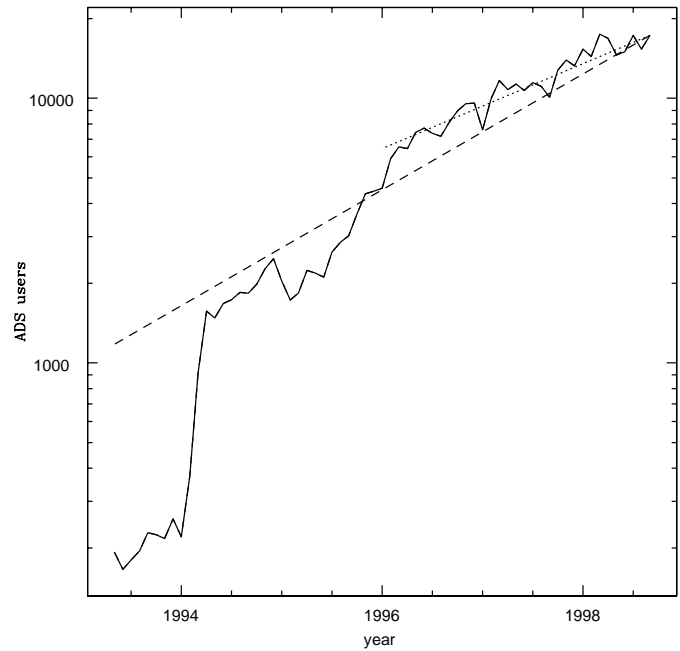}}
\caption[]{The number of users who made a query queries      made each month to the the ADS Abstract Service.  The dashed line      represents a doubling every 17 months, the dotted line a doubling      every 22 months. }
\label{num.users}
\end{figure}

Figure \ref{num.users} shows the number of unique users who made a
query using the ADS each month from April 1993 to September 1998.
Before early 1994 users had user names and passwords in the old,
proprietary system, and could be counted exactly; after the ADS became
available on the WWW users were defined as unique IP addresses.  Note
the enormous effect the WWW had on ADS use, a factor of four in the
first five weeks.  The straight dashed line represents the 17 month
doubling period seen recently in the number of queries; the dotted
line, which better represents the recent growth, is for a 22 month
doubling period.  The difference between the two is due to a one third
increase in the mean number of queries per month per user (from 19 to
25) since 1996.

>From another perspective, the number of unique IP addresses from a
single typical research site (STScI) which access the full-text data
in a typical month (September 1998) is 107, the number of unique
cookies associated with stsci.edu which access the full-text data is
104, the number of unique IP addresses from STScI which make a query
to ADS is 148 and the number of cookies is 140.  The number of AAS
members listing an STScI address is 145 (J. Johnson, personal
communication), and the number of different people listing an STScI
address in the Astropersons e-mail compilation
(\cite{1995emdw.book.....B}) is 195.  Those who access the full-text
average one article per day, those who make queries average two per
day.

We believe nearly all active astronomy researchers, as well as
students and affiliated professionals use the ADS on a regular basis.
Most of the recent exponential growth of use of the ADS is due to an
increased number of users; this growth cannot last much longer, the
17,000 who made queries in September 1998 are probably the majority of
all those who could conceivably want to make a query of the technical
astronomy literature.

\section{\label{journals} How the Astronomical Literature is used}

Electronic libraries, because they provide access to the literature on
an article basis, can provide direct measures of the use of individual
articles.  Direct bibliometric studies of article use are rare, and
tend to be based on small samples (e.g. \cite{1998JASIS..49.1283T});
most bibliometric studies use indirect measures, particularly citation
histories,(e.g. \cite{1979cita.book.....G};
\cite{1989ARIST..24..119W}; \cite{1993LibT...49..665L}), as proxies
for use.

Astronomy is perhaps unique, in that it already has an integrated
electronic information resource (ADS/Urania) which includes electronic
access to nearly all the modern journal literature, and which is used
by a large fraction of practitioners in the field, worldwide.  The
combined Urania logs, including the electronic journals and the ADS,
probably represent a fair sample of total readership in the field,
perhaps even a majority of the readership as well.

In this section we will investigate the use of the astronomy
literature as shown by the ADS logs; for articles more than a few
months past the publication date they probably represent accurately the
use of the astronomy literature.  For articles immediately after
publication the logs of the electronic journals are the definitive
source; this usage pattern is substantially different from the pattern
shown in the ADS logs, for example, the half-life for article reads for
the electronic {\it Astrophysical Journal} is measured in days
(E. Owens, 1997, personal communication).

\subsection{\label{read-use} Readership as a Function of Age}

The ADS logs provide a direct measure on the readership of individual
articles.  There are several different ADS logs, here we will use the
``data'' log.  Entries in the data log correspond to individual data
items selected from a list which is returned following a query, such
as shown in figure \ref{m87.out}.  Each entry is the result of a user,
who can see the authors and title of a paper, choosing to get more
information.  61\%\ of these requests are for the abstract, 34\%\ are
for the whole text, 2\%\ are for the citation histories, as well as
several other options; SEARCH lists all the options and their use.  In
what follows we will refer to any request for data as a ``read.''  By
``age'' we refer to the time since publication of an article, NOT the
time since birth of the astronomer reading the article!

In this subsection we restrict the study to the January 1999 log, and
only requests for information about articles published in the largest
(in terms of ADS use) eight journals ({\it ApJ, ApJL, ApJS, A\&A,
A\&AS, MNRAS, AJ, PASP}; hereafter the Big8).  The Big8 represent
62\%\ of the 270,000 entries in the January data log.

\begin{figure}
\resizebox{\hsize}{!}{\includegraphics{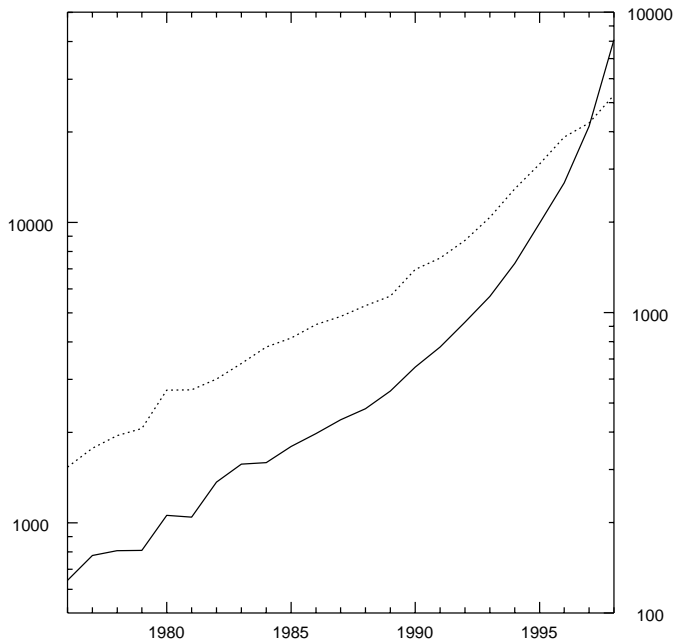}}
\caption[]{The use of journal articles via the ADS      as a function of age.  The ordinate is the publication year.  The      solid line (left abscissa) shows the total number of reads, the      dotted line (right abscissa) shows the total number of different      articles for which data was requested. }
\label{raw-use}
\end{figure}

Figure \ref{raw-use} shows the number of ADS reads (solid line, left
abscissa) during January 1999 for articles published in the Big8 from
1976 to 1998, and the number of Big8 articles for which at least one
data item was requested (dotted line, right abscissa), on a log-linear
plot, binned yearly.  The ADS database is 100\%\ complete in titles,
and in links to the full text of articles (either to the ADS scans, or
directly to the electronic journals), and is 99\%\ complete in article
abstracts for the Big8 journal articles published during this 22 year
period.

\begin{figure}
\resizebox{\hsize}{!}{\includegraphics{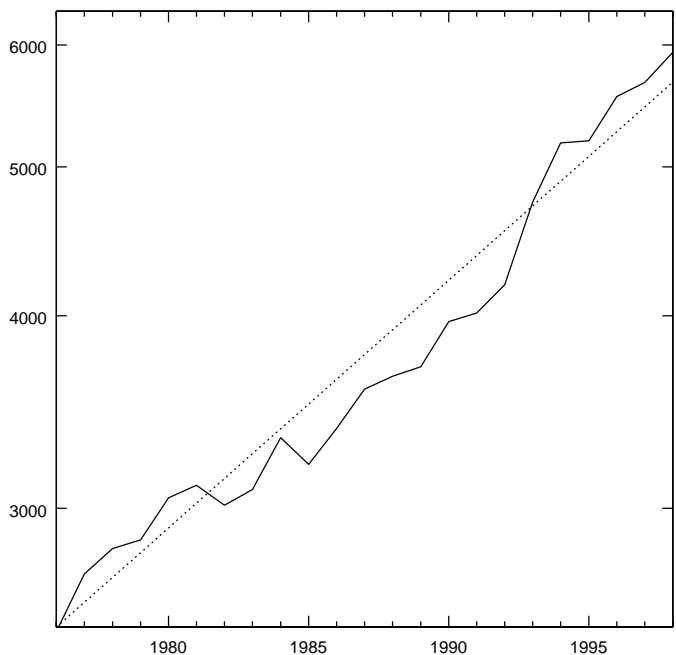}}
\caption[]{The number of Big8 journal articles published      per year.  The dotted line represents a 3.7\%\ yearly      increase. }
\label{num.pub}
\end{figure}

\begin{figure}
\resizebox{\hsize}{!}{\includegraphics{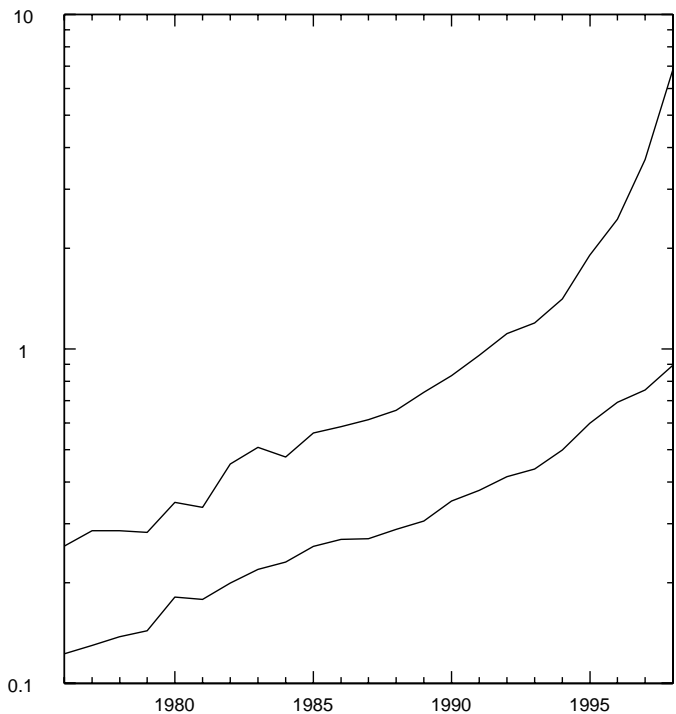}}
\caption[]{The use of journal articles via ADS as      a function of age.  The ordinate is the publication year.  The      upper line shows the mean number of reads per paper, the lower      line shows the fraction of different articles for which data was      requested.  }
\label{norm-use}
\end{figure}

The number of papers published in the Big8 has been increasing at
about 4\%\ per year during this 22 year period
(\cite{1997PASP..109.1278S}; \cite{1998PASP..110..210A}; figure
\ref{num.pub}), figure \ref{norm-use} shows the information in figure
\ref{raw-use} divided by the number of papers published.  The top line
shows the mean number of reads per paper, and the bottom line shows
the fraction (maximum 1) of papers published for which information was
requested.

>From 1976 to about 1994 the two lines are nearly parallel; this
demonstrates that the change in readership with age is caused mainly
by a change in the fraction of papers which are considered interesting
enough to be read, not by a change in the number of times an
interesting paper is read.  Extrapolating the relation seen in the
earliest 16 years of figure \ref{norm-use} we find that the fraction
of articles interesting enough to be read is $I = I_0e^{-0.075T}$,
where T is the age of the article in years, and $I_0$ is about 0.7.
Similarly readership declines as $\sim e^{-0.09T}$, so the mean
number of reads per relevant article is $M = M_0e^{-.015T}$, with $M_0$
equal to 2.5 reads per month.  For articles between 4 and 22 years old
the readership pattern is well fit by $R = IM$.

For articles younger than 4 years old the extrapolation of the $R =
IM$ model substantially underestimates readership.  While the fraction
of read papers is only about 20\%\ higher than the extrapolation (it
could not be more than 30\%\, after which all papers would be read),
the mean reads per paper is 350\%\ higher.

We postulate that there is another mode of readership, which dominates
for articles between one month and four years old, we will call this
``papers current enough to be read.''  If we subtract the $R = IM$
model from the data we get the residual of papers current enough to be
read.  This can be well represented by $C = C_0e^{-0.85T}$, where $C_0$
is equal to 5 reads per month.  Now we have a two component model for
readership (per article published), valid for papers between one month
and 22 years old which is $R = IM + C$.

\begin{figure}
\resizebox{\hsize}{!}{\includegraphics{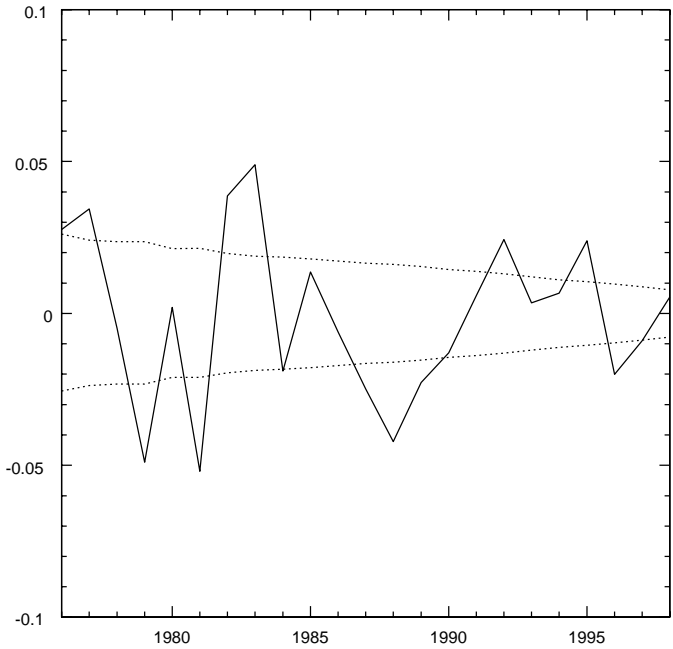}}
\caption[]{Accuracy of the $R = IM + C$ model, versus      publication date.  The abscissa is the difference between the log      of the number of reads per article published using ADS during      January 1999 and the log of the readership model described in the      text.  The dotted lines show $1 \sigma$ errors using      $\sqrt{N}$. }
\label{model-diff}
\end{figure}

Figure \ref{model-diff} shows how well the model fits the actual
readership data for January 1999.  The solid line shows the difference
between the log of the reads per paper published and the log of the
model; the dotted lines show the $1 \sigma$ errors, estimated using
$\sqrt{N}$.  Clearly the model fits the data well.

While the $R = IM + C$ model accounts for the vast majority of ADS
use, there are at least two other modes of readership, which we will
call ``historical'', and ``new''.  The historical mode describes the
use of very old articles, and the new mode describes the readership of
the current issue of a journal.

\begin{figure}
\resizebox{\hsize}{!}{\includegraphics{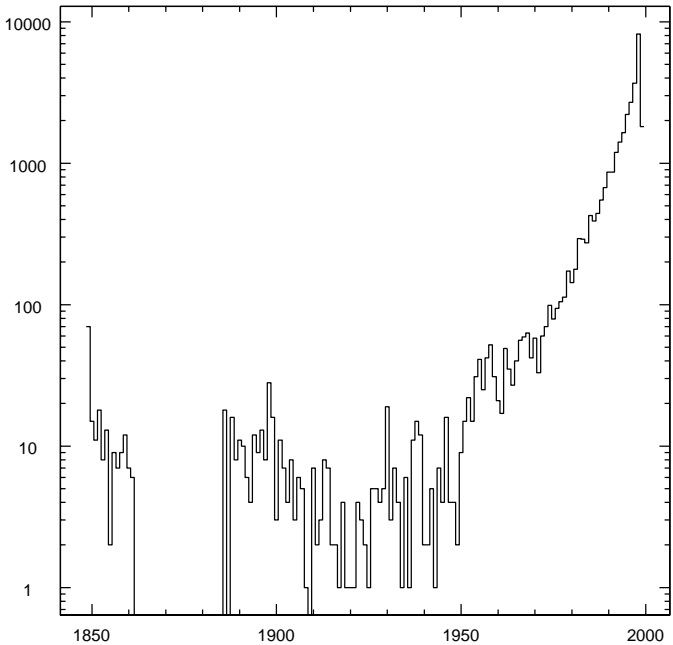}}
\caption[]{Readership of the {\it Astronomical Journal}.       Total number of reads of {\it AJ} articles using ADS during       January and February 1999, as a function of publication       year. }
\label{AJ.reads}
\end{figure}

The ADS in January 1999 had only one journal which is complete to an
early enough time to measure the historical mode, the {\it
Astronomical Journal}, which is complete from volume 1 in 1849.  The
data currently available (shown in figure \ref{AJ.reads}) suggest a
constant low level use, independent of time, $H = H_0$, where $H_0$ is
0.025 reads per month.  With the database now being extended to
include much of the literature of the past two centuries this
parameterization should improve greatly in the next couple of years.

The new mode represents the readership of the latest issue of a
journal.  As soon as a journal is issued, either received in the mail,
or posted electronically, a large number of astronomers scan the table
of contents and read the articles of interest.  Although ADS has a
feature in the Table of Contents page which supports this type of
readership, it does not represent a substantial fraction of ADS use.
We believe most users do this either with the paper copy, or through
the electronic journals directly.  We can crudely estimate this mode
in the ADS use by examining the daily usage logs following the release
of new issues of the {\it Astrophysical Journal}, After subtracting
the other modes already described we find $N = N_0e^{-16T}$, where
$N_0$ is about 3.5 reads per month.  For an accurate description of
this mode one would need to analyze the logs of the electronic
journals.

Finally we have a four component model for how the astronomical
literature is read, as a function of the age of an article, $R = N + C
+ IM + H$, where the first three terms are exponentials with very
different time constants, and the fourth is a low level constant.  ADS
use certainly underestimates the amplitude of the $N$ term, and may
underestimate the amplitude of the $C$ term, as there are alternative
electronic routes to some of these data.

\subsection{\label{cite-use} Comparison of Readership with Citation History}

Citation histories have long been used to study the long-term
readership of scientific papers (e.g. \cite{1960AmDoc..11...18B}) with
the basic result that the number of citations that a paper receives
declines exponentially with the age of the article.  While it is often
assumed that the pattern of use is similar to the pattern of citation
this has not been conclusively demonstrated.  Recently
\cite{1998JASIS..49.1283T} has found that the mean use half-life for a
set of medical journals was 3.4 years, while the mean citation
half-life for the same journals was 6.3 years.

We will compare the use of some of the Big8 journals with their
citation histories using two datasets: the ADS data logs for the
period from 1 May 1998 to 31 July 1998, and the citation information
provided to ADS by the Institute for Scientific Information covering
references in articles published during the first nine months of 1998,
and only covering references from 1981 to date.  ISI does not provide
us with the full citation histories, rather they provide us with pairs
of citing and cited journal articles where both are in the ADS
database, so the results will systematically underrepresent the
citation histories of articles with substantial influence in areas
outside astronomy, or where the primary references come from
conference proceedings.

\begin{figure}
\resizebox{\hsize}{!}{\includegraphics{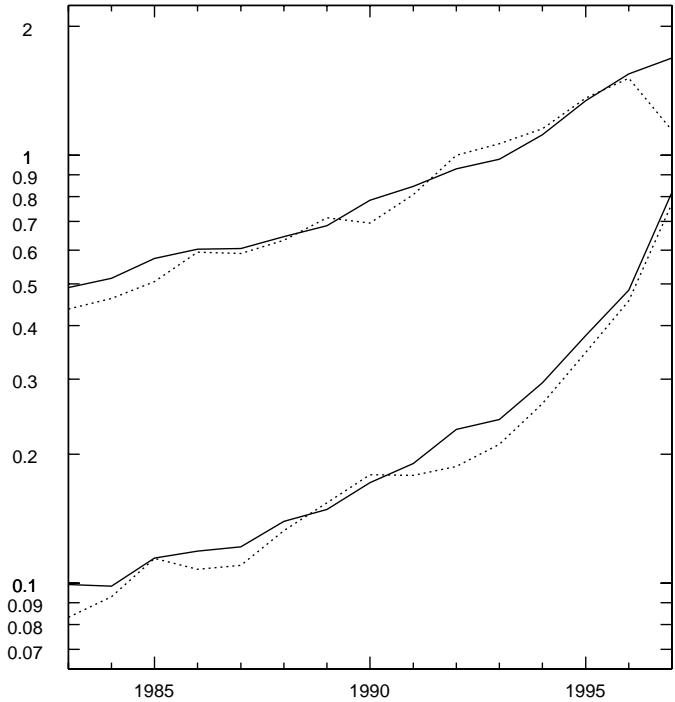}}
\caption[]{The Big8 citation rates as a function of      publication date compared with the readership rates.  The dotted      lines, and the abscissa refer to the citation information.  The      top dotted line represents the number of citations per article      for citations in papers published during the first nine months of      1998.  The bottom dotted line represents the fraction of articles      which were cited during this period.  The bottom solid line shows      the number of reads per article for the three month period      May-July 1998, and the top solid line shows the fraction of      articles read.  Both solid lines are arbitrarily shifted to show      the similarity of the functional shapes. }
\label{reads-cites}
\end{figure}

Figure \ref{reads-cites} compares the citation histories of the Big8 
journals with their readership; the abscissa refers to the citation 
information (dotted lines), the readership data (solid lines) have been
arbitrarily shifted for comparison.  The lower dotted line represents the 
fraction of Big8 journal articles which were cited during the first nine
months of 1998; the upper dotted line represents the mean number of cites 
per article.  The lower solid line shows the mean number of reads per
article during the three month period May-July 1998, shifted by a factor
of 19; the upper solid line shows the fraction of Big8 articles read, times
1.8.

The number of cites has the same functional form as the fraction of
reads, And the fraction of cites has the same form as the number of
reads.  This result is perhaps surprising.

Except for the most recent year (1997), where the number of cites
declined from the year before the number of cites per article declines
with age as $\sim e^{-0.09T}$ or proportional to $IM$, the long term
declining readership.  The citation half-life for these articles, 7.7
years, is longer than the 4.9 years found by \cite{1990JASIS..41..283G} for
the {\it Physical Review}, but is consistent with results of Abt
(1981, 1996) of 20-30 year half-lives with no normalization, once one
takes the increase in the number of astronomy papers/cites into
account (Abt 1981, 1995).

The fraction of articles cited, on the other hand, appears to follow
the same two component form as readership, $R = IM + C$.  We postulate
the following explanation for this behavior.  The degree of citability
we define as the degree to which a paper would be cited, were it
possible.  We postulate this is directly proportional to readership:
$D = D_0R$.  The large increase in the fraction of recent papers cited
is thus due to the large increase in readership.  We define the
ability of a paper to be cited to be a steeply increasing function of
age, simply because for one paper to cite another it must appear
before the second paper is written, refereed, and published: $A = 1 -
e^{-1.5T}$.  Our model for the mean number of citations a paper
receives, $Z$, as a function of age is: $Z = Z_0AD$ or $Z = Z_0AD_0R$.

\begin{figure}
\resizebox{\hsize}{!}{\includegraphics{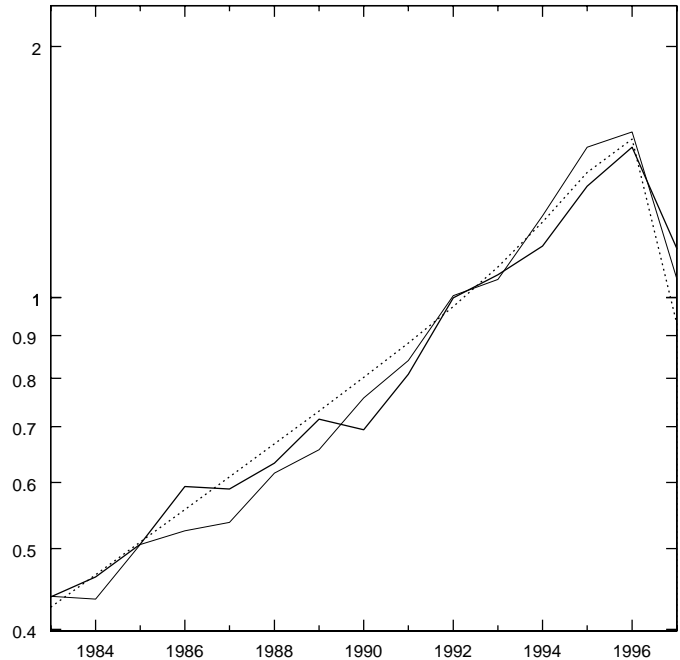}}
\caption[]{The number of citations per article versus      the $Z = 0.08R(1 - e^{-1.5T})$ model.  The thick solid line      represents the number of citations per article from papers      published in the first nine months of 1998, as a function of      publication date.  The thin solid line represents the model,      where $R$ is the actual readership data; the dotted line      represents the model, where $R$ is the $R = IM + C$      model. }
\label{cites-fit}
\end{figure}

Figure \ref{cites-fit} shows the the number of citations per paper as
a function of age (thick solid line), the $Z = Z_0AD_0R$ model using
the actual number of reads per paper for $R$ (thin solid line), and
the $Z = Z_0AD_0R$ model using the $R = IM + C$ model for $R$ (dotted
line).  The product of the constants $Z_0D_0$ is the number of
citations per read, currently this is about 0.08.

The papers which are frequently cited tend also to be frequently read,
although the correlation is not very strong.  We rank the papers by
number of cites/reads during the 1998 periods, and perform a Spearman
rank correlation between the 26988 different Big8 papers cited and the
53755 papers read (57340 total), we obtain $r_{Spearman} = 0.35$.
This underestimates the correlation because it excludes papers which
were neither cited nor read.

\begin{figure}
\resizebox{\hsize}{!}{\includegraphics{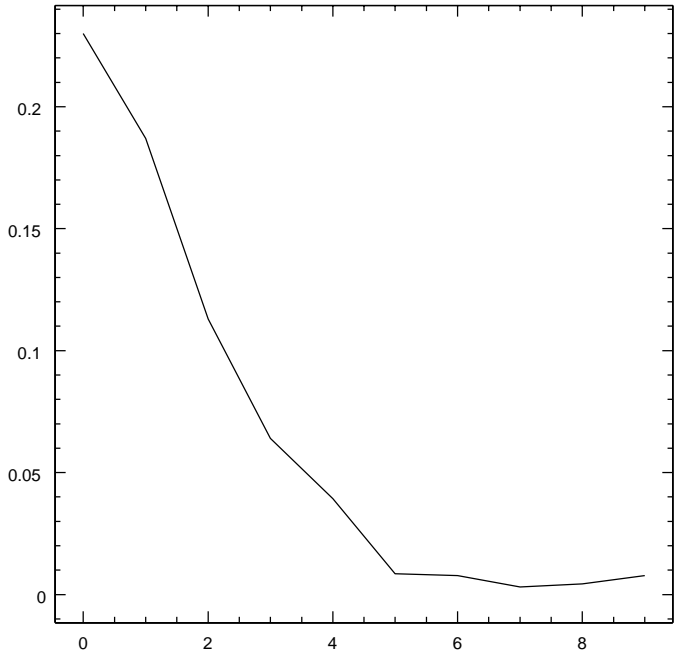}}
\caption[]{Fraction of Big8 papers unread during a 3      month period in 1998, as a function of the number of times the      papers were cited during a nine month period in      1998. }
\label{frac-unread}
\end{figure}

Of the 66392 Big8 papers published between 1982 and 1997 81\%\ were
read in the 3 month period using ADS, while 41\%\ were cited during
the 9 month period.  The probability that a paper was not read
declined sharply with the number of times it was cited.  Figure
\ref{frac-unread} shows this; one paper each of the (324, 224, 126)
papers which were cited (7, 8, 9) times went unread during the period;
none of the 430 papers which were cited 10 or more times went unread.

The relations between the number of cites or reads of a paper and the
rank that paper has when ranked by number of cites/reads are
identical.  If one takes papers published in a single year both cites
and reads follow a \cite{1949hbplebook.....Z} power law $n \sim
r^{-\alpha}$ ($n$ is the number of reads or cites, and $r$ is the rank
of the paper with that many reads/cites), where $\alpha$ is
${1}\over{2}$, this is the same result \cite{EPhJB...4..131R} found
for citation histories for the physics literature.  If papers from all
years are taken together and ranked the power law index flattens
identically for both cites and reads to $\alpha = {{1}\over{3}}$.

\subsection{\label{journal-use} How the Journals are Used}

\subsubsection{\label{main} The main journals}

\begin{figure}
\resizebox{\hsize}{!}{\includegraphics{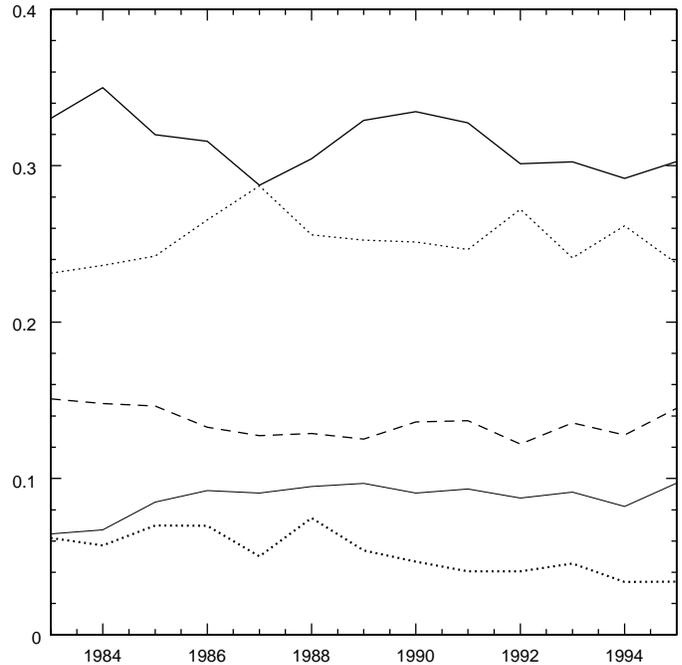}}
\caption[]{Fraction of Big8 papers published by five      selected journals.  The top line (thick. solid) is {\it ApJ},      below that (dotted) is {\it A\&A}, in the middle (dashed) is {\it      MNRAS}, second from the bottom (thin, solid) is {\it AJ}, and the      lowest line (thick, dotted) represents {\it PASP}.      }
\label{frac.pub}
\end{figure}

\begin{figure}
\resizebox{\hsize}{!}{\includegraphics{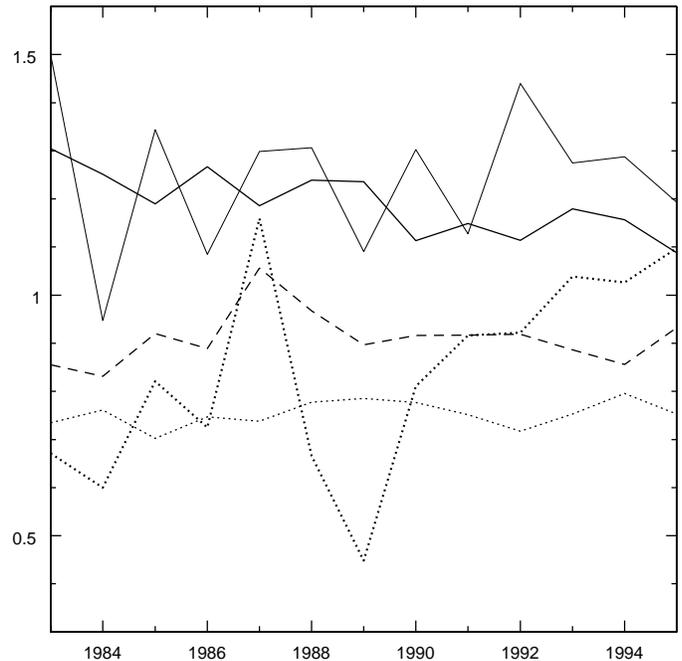}}
\caption[]{Readership rates for five journals.      Linetypes are as in figure \ref{frac.pub}.  The lines represent      the ratio of the fraction of reads of articles in a given journal      to the fraction of articles that journal published.  Note that      the large spike for {\it PASP} in 1987 is due to a single very      well read paper \cite{1987PASP...99..191S} combined with      fluctuations in the number of conference proceeding abstracts      published in the journal. }
\label{norm-reads}
\end{figure}

\begin{figure}
\resizebox{\hsize}{!}{\includegraphics{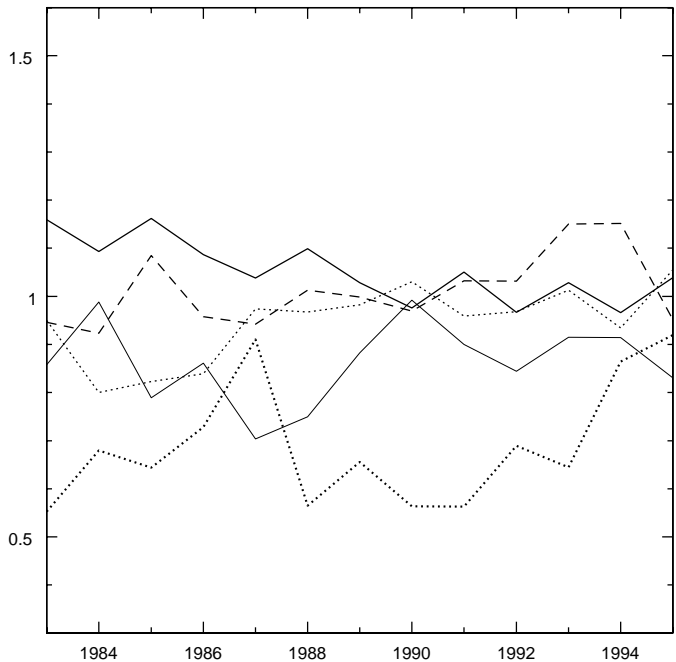}}
\caption[]{Citation rates for five journals.      Linetypes are as in figure \ref{frac.pub}.  The lines represent      the ratio of the fraction of cites of articles in a given journal      to the fraction of articles read in journal.  Note that the large      spike for {\it PASP} in 1987 is again due to a single very well      cited paper \cite{1987PASP...99..191S}. }
\label{norm-cites}
\end{figure}

Figure \ref{frac.pub} shows the fraction of articles published in the
Big8 by each of the five main journals, leaving out the letters and
supplements. We show the data only for articles published from 1983 to
1995.  Before 1983 the data from ISI are less complete, and after 1995
the presence of the electronic journals, and the differing rules for
the distribution of the ADS bitmaps, make the meaning of a ``read''
differ from journal to journal.  The reads and cites data for figures
\ref{frac.pub}, \ref{norm-reads}, and \ref{norm-cites} comes from the
same 1998 reporting periods described above.

Figure \ref{norm-reads} shows the relative readership of papers as a
function of journal and publication year.  The abscissa is the ratio
of the fraction of Big8 papers read and the fraction of Big8 papers
published.  Were all papers read equally frequently, independent on
the journal in which they were published, figure \ref{norm-reads}
would show five straight lines at one; it does not.  The papers from
the {\it AJ} are read more on a per article basis than the other
journals; the papers from {\it A\&A} are read less.  Recent {\it PASP}
papers are read substantially more frequently than older ones, when
compared with the readership patterns of the other journals.

Figure \ref{norm-cites} shows the ratio of the fraction of citations
an article received to the fraction of reads, as a function of journal
and year.  Were all articles cited in the same proportion to the
number of times they were read (this is the constant $Z_0D_0$ in
\ref{cite-use}) then the figure would be five straight lines at one.
The three bi- and tri-monthly journals do not show much deviation from
straight lines at one, while the {\it AJ} appears to be systematically
less cited than it is read.  The {\it PASP} again shows an increase
during the beginning of this decade.

Recall that the readership and citation information are from hundreds
of thousands of individual decisions made by more than 10,000
astronomers during 1998.  Taken together figures \ref{frac.pub},
\ref{norm-reads}, and \ref{norm-cites} show the current opinion of
astronomers as to the usefulness of articles as a function of journal
and publication date.  The growth of the {\it AJ} for example, from
6.5\%\ of Big8 articles to 9.5\%\ has not greatly affected the
relative readership or citation rates for the journal.

The recent history of the {\it PASP} is perhaps the most interesting
feature on figures \ref{frac.pub}, \ref{norm-reads}, and
\ref{norm-cites}.  From 1983 to 1995 the fraction of Big8 papers
published by {\it PASP} declined from 6\%\ to 3\%\ .  This decline is
overstated, as {\it PASP} published some conference proceeding
abstracts during the late 1980s, a practice which ended in 1991; the
decline is nevertheless real: {\it PASP} published the same number of
papers in 1995 as 19 years before, during which time the number of
Big8 journal articles doubled.

Figure \ref{norm-reads} shows two main features, fluctuations, and a
slow rise.  The large fluctuations during the late 80s and early 90s
are due to two factors: fluctuations in the number of conference
proceeding papers and abstracts; and the influence of
\cite{1987PASP...99..191S}, which was read at twice the rate of the
next most read paper from 1997, and four times the next most read {\it
PASP} paper from that year.  The rise in the readership measure during
the 1990s is not caused by any known systematic; we believe it
represents a real increase in the perceived usefulness of the journal.

Figure \ref{norm-cites} also shows the influence of
\cite{1987PASP...99..191S}, currently the third most cited article in
the ADS database, although now without the addition of the
fluctuations in article counts.  It also shows the rise in the
perceived usefulness per article (this time in the measure of cites
per read).  Noting that the number of cites per article is the product
of figures \ref{norm-reads} and \ref{norm-cites} the rise in the
number of cites per article, compared with the Big8 over the period
1989 to 1995 is a factor of three, so that now the journal is at full
parity with the Big8.  This demonstrates that the policy during this
period was one of quality rather than quantity, a policy we dub
``shaken, not stirred.''

\subsubsection{\label{currency} Loss of relative currency}

All Big8 astronomical journals lose currency, the current usefulness
of an article, at a rate described by the readership and citation
models of \ref{read-use} and \ref{cite-use}.  Any changes in the loss
of currency of one journal with respect to the rest of the Big8 should
be seen in figure \ref{norm-reads} in the form of a relative decrease
in readership, as a function of age.  Indeed the changes in the {\it
PASP} which we have attributed to changes in editorial policy could
simply be a substantial loss of relative currency.

\begin{figure}
\resizebox{\hsize}{!}{\includegraphics{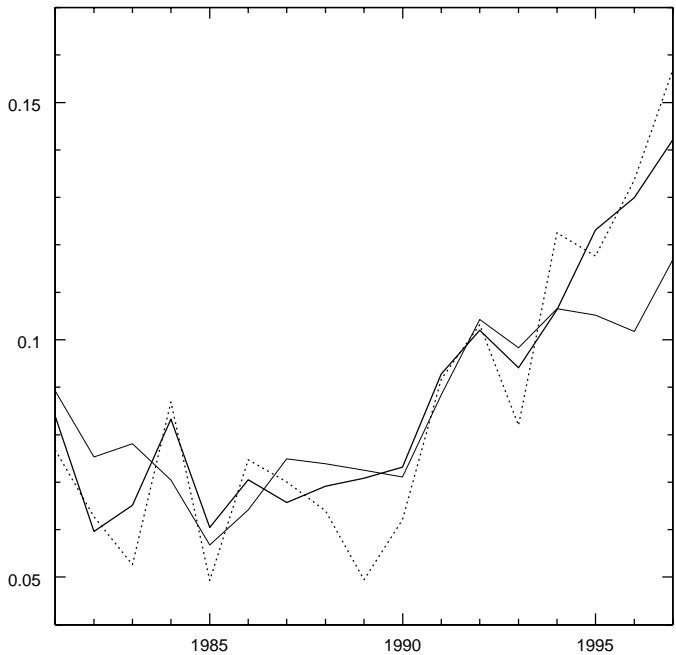}}
\caption[]{Use of the {\it Astrophysical Journal       Letters} from 1981 to 1997.  The thin solid line shows the       fraction of Big8 papers published in {\it ApJL}, the thick solid       line the fraction of reads, and the dotted line the fraction of       cites. }
\label{apjl.use}
\end{figure}

One of the Big8 journals, the {\it Astrophysical Journal Letters} is
intended to lose currency more rapidly than the other journals.
Figure \ref{apjl.use} shows the relative fraction of articles
published (thin solid), articles read (thick solid), and articles
cited (dotted) for the {\it ApJL} from 1981 to 1997.  Except for the
period from 1994 to 1997 the curves track each other reasonably well;
older {\it ApJL} papers are not cited or read any more or less than
the Big8 average.  For the more recent papers the cites and reads
increase above the fraction published, implying that the journal is in
some sense more current than average.

In terms of readership this effect is strongly affected by a
systematic.  During the 3 month period in 1998, most of the 1996 and
all of the 1997 issues of {\it MNRAS} were not available
electronically due to copyright constraints.  This dramatically
lowered the relative readership of that journal, pushing all the
others up.  Also all five journals which were fully electronic during
1997 show increases compared with {\it AJ} and {\it PASP} which were
only available as bitmaps.  Thus the increase in readership of the
{\it ApJL}, the pioneer electronic journal (\cite{1995AAS...187.3801B}),
could be due to its superior delivery system, rather than its content.

\subsubsection{\label{local-diff} Local differences in readership rates}

\begin{figure}
\resizebox{\hsize}{!}{\includegraphics{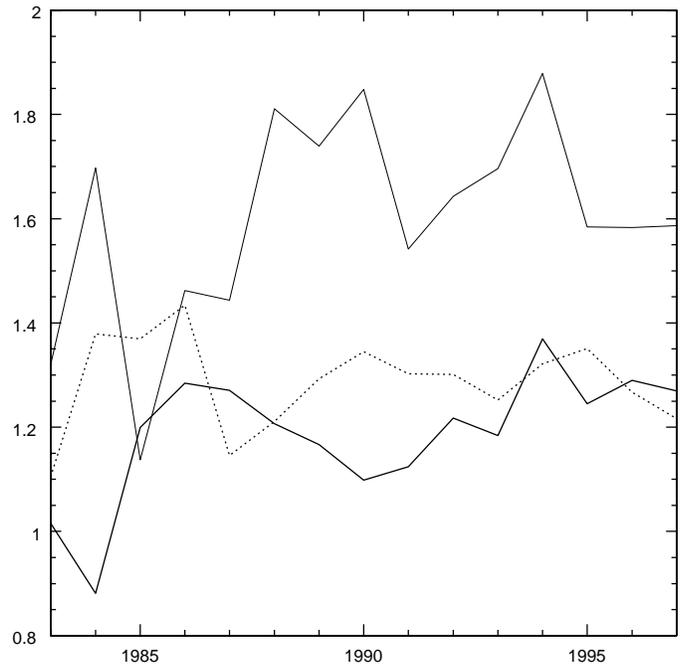}}
\caption[]{Local differences in readership rates       for three journals.  The thin solid line shows the increased use       of {\it MNRAS} in the UK compared with the rest of the world;       the dotted line shows this for {\it A\&A} in Europe, and the       thick solid line for {\it AJ} in the US.       }
\label{uk-fr-us.ratio}
\end{figure}

Astronomers in different parts of the world read different journals at
different rates than the average.  Figure \ref{uk-fr-us.ratio} shows
three typical differences.  The three curves show the ratio of
readership fractions for a particular subset when compared with the
rest of the world; a value of 1 means that there is no difference in
relative readership.  The thin solid line shows the {\it MNRAS}
readership ratio for users who access the US site and have IP
addresses ending in .uk; it shows that the British read {\it Monthly
Notices} about 60\%\ more than the world average.

The dotted line shows the {\it A\&A} readership ratio for users of the
Strasbourg mirror, and the thick solid line shows the {\it AJ}
readership ratio for US users with an IP address ending in .edu.  They
show that Europeans/Americans read {\it A\&A}/{\it AJ} about 20\%\
more than the rest of the world.  The {\it ApJ} also shows about a
20\%\ increase in the US; the {\it PASJ} shows a 300\%\ increase in
Japan.

\subsubsection{\label{AJhistory} Use of historical literature}

The ADS is in the process of putting a large fraction of the
astronomical literature of the past two centuries on-line via
bitmapped scans.  The first nineteenth century journal to be fully
on-line is the {\it Astronomical Journal}, which was first fully
on-line on 1 January 1999.  Figure \ref{AJ.reads} shows the raw readership
figures for the first two months of 1999 (US logs only), this shows
the current readership of 150 years of the journal.

Clearly the back issues are being read; the only year where the
journal was published, but no paper was read in the two months, was
1909, where only 12 papers were published.  Also there is a break in
the exponential falloff with age for articles published between 1950
and 1960, where approximately twice the expected readership occurred.
During this period 94 different users read 283 articles; the biggest
user made 13 reads.  We have no explanation for this increased use.
The only other period where the use is not predicted by the $C + IM +
H$ model of \ref{read-use} is the first decade of the journal's
existence, perhaps due to curiosity.

\section{\label{impact} The Impact of the ADS on Astronomy}

It is difficult to judge the impact of scientific work.  For
scientific programs citation histories, personal honors and awards,
and the success of students can give a measure of impact.  For support
type programs these measures do not suffice; the impact of the
200-inch Hale Telescope (\cite{1948PASP...60..221A},
\cite{1948PASP...60..225R}) or the 4-meter Mayall Telescope
(\cite{1965S+T....29..268C}) clearly extends beyond the papers and
honors of their respective developers.  The impact of large software
projects is, if anything, even harder to quantify; the large data
reduction environments, like AIPS (\cite{1981NRAON...3....3F},
\cite{1998aipsm.100....1G}), MIDAS (\cite{1983Msngr..31...26B}), or
IRAF (\cite{1986SPIE..627..733T}) have transformed astronomy, but how
much?

The ADS is perhaps unique among large support projects in that a
reasonably accurate quantitative estimate of its impact can be made.
This is because many of the services the ADS provides are just more
efficient methods of doing things astronomers have long done, and
found worth the time it took to do them.

We will assign to each of several ADS functions a time which is our
estimate of the increase in research time which accrues to the
researcher by virtue of using that function.  Our fundamental measure
will be the time saved in obtaining an article via the ADS, which we
estimate from the time it takes to go to the library, find the volume,
photocopy the article, and return to the office, as 15 minutes.  We
then estimate that reading an abstract, a reference list, or a
citation history saves $1/3$ of the full article time, or 5 minutes,
and we arbitrarily assign a one minute time savings to each query.

We can now estimate the impact of ADS, in terms of FTE (Full Time
Equivalent, 2000 hour) research years, by examining the ADS usage
logs.  We note that about half of the full text articles currently
retrieved via the ADS come from the on-line journals, which certainly
deserve credit for their work.  Also we are ignoring several important
(but hard to quantify) aspects of the ADS service, such as links from
other web sites (e.g. the HTML journals), the synergy of joint
ADS/SIMBAD and ADS/NED queries (e.g. that in figure \ref{m87.query}),
the bulk retrieval of abstracts and LATEX formatted references (about
200,000 per month), and the more than 10,000,000 references returned
each month.  We think that what follows is a reasonable estimate of
the impact of the ADS on astronomy, and that the impact of the full
Urania collaboration is substantially more.

Using the March 1999 worldwide combined ADS logs there were 113,471
full text articles retrieved, 195,026 abstracts (individually
selected), 10,663 citation histories, and 3,702 reference pages
retrieved, and 582,836 queries made.  Using the estimated time savings
above we find that the impact of the ADS on astronomy is 333 FTE
research years per year, approximately the same as the entire
Harvard-Smithsonian Center for Astrophysics.

If we crudely estimate that there are 10,000 FTE research years in
astronomy each year the ADS can be viewed as accounting for 3.33\%\ of
astronomy.  Currently the ADS contains 27,712 (11,834) articles
(refereed articles) in the astronomy database dated 1998, so one way
of expressing the impact of the ADS would be 923 (394) articles
(refereed articles) per year.

While the efficiencies brought about by the technologies inherent in
the ADS and Urania are permanent, and will contribute (compounded) to
the accelerating pace of discovery in astronomy, one can ask what was
gained by being first.  Risks were taken in funding the early
development and adoption of technologies via the ADS and Urania.
Also, had nothing been done, the ``winning'' technologies would
eventually be adopted with very little risk.

To judge the payoff we adopt a simple model; we assume that the
increase in research efficiency due to the ADS has increased linearly
from zero in 1993 to 333 FTE research years in 1999, and that it will
decrease linearly to zero over the next six years, after which there
will be no difference in the technologies employed.

This yields a sum impact from the early creation of the ADS of 2,332
FTE research years, which is 23\%\ of the astronomical research done
in a single year, or 6463 (2760) papers (refereed papers).  This is
surely equal to the impact of the very largest and most successful
projects.  Doing this analysis for the entire Urania would yield a
substantially increased amount.

\section{\label{acknowldgements} Acknowledgments}

Peter Ossorio is a pioneer in the field of automated text retrieval,
he gave freely of his ideas in the early phase of the project.  Geoff
Shaw provided the enthusiasm to keep the Abstract Service project going
during the long period of no funding.

Margaret Geller gave crucial encouragement at the time of the original
prototype. Frank Giovane long believed in the possibilities of the
Abstract Service, and acted as a friend in high places.

Todd Karakashian wrote much of the software at the time of the public
release, he left in 1994.  Markus Demleitner joined the ADS project in
April 1999, he has already produced much of value.

There are about a dozen individuals at the Strasbourg Observatory, and
the Strasbourg Data Center to thank, too many to thank individually.
The data services provided by them are at the heart of the new
astronomy; their collaboration with the ADS has been both very fruitful,
and a great joy.

Peter Boyce, Evan Owens, and the electronic Astrophysical Journal
project staff have had the vision necessary to do things first.  Their
collaboration has been important to the success of the ADS, and
crucial to the success of Urania.

Without the long term support from NASA, and G\"unter Riegler in
particular, the ADS would not now exist.

We are supported by NASA under Grant NCC5-189.

\end{document}